\documentclass[11pt,preprint,a4paper,prd,superscriptaddress]{revtex4-2}

\usepackage{amsmath,amssymb}
\usepackage{graphicx,}

\usepackage{mathtools}
\usepackage{physics,}

\usepackage[pdftex,dvipsnames,hyperref]{xcolor} 
\usepackage[pdftex,pdfencoding=auto,psdextra,colorlinks=true,allcolors = blue]{hyperref}

\definecolor{red}{RGB}{255,75,0}    %red in the CUD
\definecolor{green}{RGB}{3,175,122} %green in the CUD
\definecolor{blue}{RGB}{0,90,255}   %blue in the CUD

\DeclareMathOperator{\corank}{corank}
\DeclareMathOperator{\codim}{codim}
\DeclareMathOperator{\Span}{Span}

\newcommand{{\COne}}{C1}

\newcommand{\HypersurfaceDimension}{D}
\newcommand{\SpacetimeDimension}{\HypersurfaceDimension + 1}
\newcommand{\Spacetime}{\mathcal{M}}
\newcommand{\Worldsheet}{\mathcal{S}}
\newcommand{\TmpOneForm}{A}

\newcommand{\KillingVector}{k}
\newcommand{\KillingOneform}{\eta}
\newcommand{\KillingTwoform}{\dd{\KillingOneform}}
\newcommand{\Submanifold}{\mathcal{W}}

\newcommand{\OneParamGroupIsom}{\varphi_v}
\newcommand{\Hypersurface}{\Sigma}
\newcommand{\Orbit}{\mathcal{O}}
\newcommand{\Point}{p}          %point on \Hypersurface_0
\newcommand{\Oneform}{\eta}     % 1-form on the \Hypersurface_0,
\newcommand{\Curve}{\mathcal{C}}

\newcommand{\AlmostContactStructure}{(\Endmorphism, \VectorField, \Oneform)}
\newcommand{\VectorField}{\xi}
\newcommand{\Endmorphism}{\varphi}

\newcommand{\Rank}{r}
\newcommand{\Corank}{s}
\newcommand{\Metric}{h}

\newcommand{\Codimension}{\codim\Worldsheet}

\newcommand{\IntOdd}{n}         %"n" for the odd integer 2 n + 1

\newcommand{\Ppwaves}{{\textit{pp}\,-waves}}
\newcommand{\AlmostContactMetricStructure}{(\Endmorphism, \VectorField, \Oneform, \Metric)}
\newcommand{\nijenhuis}{\mathcal{T}}

\newcommand{\SecFundForm}{\alpha}
\newcommand{\SetVecField}{\mathfrak{X}}
\newcommand{\SetFunc}{\mathfrak{F}}

\newcommand{\TwistPotential}{\omega}

\begin{document}

\title{%
  Nambu-Goto Strings with a null symmetry
  and 
  contact structure}
\author{Hiroshi \surname{Kozaki}}
\email{kozaki@ishikawa-nct.ac.jp}
\affiliation{Department of General Education, National Institute of
  Technology, Ishikawa College, Ishikawa 929-0392, Japan}

\author{Tatsuhiko Koike}
\email{koike@phys.keio.ac.jp}
\affiliation{Department of Physics, Keio University, Yokohama 223-8522,
  Japan} 
\affiliation{Quantum Computing Center, Keio University, Yokohama 223-8522,
  Japan}
\affiliation{Research and Education Center for Natural Sciences,
  Keio University, Yokohama 223-8521, Japan}

\author{Yoshiyuki Morisawa}
\email{morisawa@omu.ac.jp}
\affiliation{Osaka Central Advanced Mathematical Institute, Osaka Metropolitan University,
  Osaka 558-8585, Japan}

\author{Hideki Ishihara}
\email{h.ishihara@omu.ac.jp}
\affiliation{Nambu Yoichiro Institute of Theoretical and Experimental
  Physics, Osaka Metropolitan University, Osaka 558-8585, Japan}
\affiliation{Osaka Central Advanced Mathematical Institute, Osaka Metropolitan University,
  Osaka 558-8585, Japan}

\preprint{OCU-PHYS-577}
\preprint{AP-GR-188}
\preprint{NITEP-154}
\begin{abstract}
   We study the classical dynamics of the Nambu-Goto strings with a null
   symmetry in curved spacetimes admitting a null Killing vector field. 
   The Nambu-Goto equation is reduced to first-order ordinary
   differential equations and is always integrable in contrast to the
   case of non-null symmetries where integrability requires additional
   spacetime symmetries. 
   It is found that in the case of null symmetry,
   an almost contact structure associated with the metric dual $1$-form
   $\KillingOneform$ of the null Killing vector field emerges naturally.
   This structure determines the allowed class of string
   worldsheets in such a way that the tangent vector fields of the
   worldsheet lie in $\ker{\KillingTwoform}$. 
   In the special case that the almost contact structure becomes a
   contact structure, 
   its Reeb vector field  completely characterizes the worldsheet.
   We apply our formulation to the strings in the {\Ppwaves}, the Einstein
   static universe and the G\"{o}del universe.
   We also study their worldsheet geometry in detail.
\end{abstract}

\maketitle

\section{Introduction}

Strings, one-dimensional objects, appear in various areas of physics.
In cosmology, one-dimensional topological defects called the cosmic strings 
are supposed to have formed in the early universe
(e.g. \cite{Vilenkin:2000jqa}).  
In string theories, microscopic strings are considered to be the fundamental
elements (e.g. \cite{polchinski_1998}).

The string dynamics is characterized by a two-dimensional worldsheet 
in spacetime and is in most cases governed by the Nambu-Goto action.
The equation of motion is given by a set of partial differential equations,
and is therefore generally difficult to solve.
However, a simplification occurs when the string worldsheet has a symmetry
called cohomogeneity one ({\COne}) \cite{Ishihara:2005nu}.
The {\COne} symmetry means that the string worldsheet shares a Killing vector
field with the spacetime, or more precisely, a Killing vector field of the
spacetime is tangent to the worldsheet.

String dynamics with {\COne} symmetry have been widely studied in various
contexts. 
One example is the stationary strings.
They move in stationary spacetimes and sweep the worldsheets tangent to the
timelike Killing vector fields.
The stationary (rotating) strings in black hole spacetimes have been
studied extensively with astrophysical and geometrical interests
\cite{Frolov:1988zn,Carter:1989bs,Kinoshita:2016lqd,Boos:2017qbx,Igata:2018kry}.
Another example can be found in the context of the AdS/CFT correspondence.
The {\COne} string ansatz is effectively used in curved backgrounds,
and then the string motion is found to be chaotic
\cite{PandoZayas:2010xpn, Basu:2011dg, Ishii:2016rlk, Basu:2011di,
  Basu:2011fw, Asano:2015qwa, Rigatos:2020hlq}.
The classification problem of {\COne} strings are also studied in some
highly symmetric spacetimes such as
Minkowski spacetime \cite{Ishihara:2005nu},
five-dimensional anti-de Sitter spacetime $\text{AdS}^5$ \cite{Koike:2008fs}
and higher-dimensional flat spacetimes $\mathbb{R}^{n,1}$
\cite{Ida:2020lqk}.
The concept of the cohomogeneity one ({\COne}) symmetry has been  extended to
higher dimensional objects such as membranes
\cite{Kozaki:2014aaa,Hasegawa:2018cyk}.

In previous studies of {\COne} strings,
the Killing vector field tangent to the worldsheet is assumed to be timelike
or spacelike. 
In this case,
the Nambu-Goto equation of motion is reduced to the geodesic equation with
respect to a certain metric weighted by the squared norm of 
the Killing vector field \cite{Ishihara:2005nu,Morisawa:2019qmp}.
If the metric admits a sufficient number of Killing vector fields and
Killing tensor fields,
the geodesic equation admits a sufficient number of conserved quantities
and is then integrable in quadrature.
Indeed, it has been clarified that the {\COne} string dynamics is integrable
in some highly symmetric spacetimes
\cite{Kozaki:2009jj,% Koike:2008fs
  Morisawa:2019qmp,Ida:2020lqk}.  
On the other hand, {\COne} strings with null tangent Killing vector fields  
have not been well studied.
Strings with a null symmetry may be interesting,
for example, in {\Ppwaves}, which attract much attention in string theories
\cite{Amati:1988ww,Horowitz:1989bv,deVega:1990nr,deVega:1990kk}.

The purpose of this paper is to formulate the dynamics of Nambu-Goto strings
with a null {\COne} symmetry in curved spacetimes
and to study the dynamics, in particular, the integrability and the
extrinsic geometry of the worldsheets.

We will see that the Nambu-Goto equation reduces to ordinary differential
equations (ODEs). 
While the ODEs in the case of non-null {\COne} symmetry are second order,
the null {\COne} ODEs are first order.
The Nambu-Goto equation is always integrable in the null case, 
in contrast to the non-null case where integrability requires additional
symmetries.

We will also find that an almost contact structure associated with the
metric dual $1$-form of the null Killing vector field emerges naturally.
In the special case that the almost contact structure becomes the contact
structure, its Reeb vector field completely characterizes the worldsheet.

Contact structures appear in various areas of physics: for example,
classical dynamics \cite{arnol2013mathematical}, thermodynamics
\cite{MRUGALA1991109,Rajeev:2007uk} and electromagnetism \cite{Dahl:2004}.
Contact and almost contact structures are
lower level structures of the Sasaki structure
\cite{Sasaki:1960,Blair:1976},
which is attracting renewed attention in the context of the AdS/CFT
correspondence \cite{Gauntlett:2004yd}. 
Three-dimensional Sasaki or quasi-Sasaki manifolds are effectively used
to construct G\"{o}del-type solutions in Einstein-Maxwell-scalar field
theories \cite{Ishihara:2021abn} and a generalized Einstein's static universe
\cite{Ishihara:2021gty}.
Then, our results suggest that the lower level structures such as (almost)
contact structure may also be useful in general relativity as well.

The paper is organized as follows.
In the following section, we reduce the Nambu-Goto equation and gauge
conditions to first-order ordinary differential equations.
In Sec.~\ref{sec:general_solutions}, we solve the equations in general,
and then discuss the relation with the (almost) contact structure. 
In Sec.~\ref{sec:extrinsic_geometry}, we study the extrinsic geometry of 
the worldsheet, in particular, the second fundamental form.
In Sec.~\ref{sec:examples}, we apply our formulation to the strings in the
{\Ppwaves}, the Einstein static universe and the G\"{o}del universe
and investigate their worldsheet geometry.
Sec.~\ref{sec:conclusion} is devoted to conclusions.

\section{Equation of motion}
\label{sec:equations_of_motion}

\subsection{Equation of motion in double null coordinates}
\label{subsec:eom_double_null}
Let $(\Spacetime,g)$ be a $(\SpacetimeDimension)$-dimensional spacetime
furnished with a Lorentzian metric $g$. A string sweeps the so-called
worldsheet $\Worldsheet$, which is a two-dimensional timelike surface
\begin{align}
  x^\mu
  =
  x^\mu(\zeta^1,\zeta^2)
  \quad
  (\mu = 0, \dots, \HypersurfaceDimension)
\end{align}
where $x^{\mu}$ are spacetime coordinates or embedding functions of the
worldsheet and $\zeta^a~(a = 1,2)$ are worldsheet coordinates.
We assume that the string dynamics is governed by the Nambu-Goto action
\begin{align}
  S
  =
  \int \sqrt{-\gamma} \,\dd{\zeta^1} \dd{\zeta^2},
  \quad
  \gamma \coloneqq \det \gamma_{ab},
\end{align}
where $\gamma_{ab}$ is the worldsheet metric given by
\begin{align}
  \gamma_{ab}
  =
  g_{\mu\nu} \pdv{x^\mu}{\zeta^a} \pdv{x^\nu}{\zeta^b}.
\end{align}
Varying the action, we obtain the equations of motion
\begin{align}
  \pdv{}{\zeta^a}
  \qty(
  \sqrt{-\gamma} \, \gamma^{ab}
  \pdv{x^{\mu}}{\zeta^b}
  )
  +
  \sqrt{-\gamma} \, \gamma^{ab} \Gamma^{\mu}{}_{\nu\lambda}
  \pdv{x^{\nu}}{\zeta^a}
  \pdv{x^{\lambda}}{\zeta^b}
  =
  0,
  \label{eq:NambuGotoEqGeneral}
\end{align}
where $\Gamma^{\mu}{}_{\nu\lambda}$ is the Christoffel symbol.

In this paper, we take both of the worldsheet coordinates $\zeta^1, \zeta^2$
to be null. Then the worldsheet metric has a cross term only:
\begin{align}
  \dd{s}^2
  =
  \gamma_{ab} \dd{\zeta^a} \dd{\zeta^b}
  =
  2 \, \gamma_{12}(\zeta^1,\zeta^2) \dd{\zeta^1} \dd{\zeta^2},
  \quad
  \gamma_{12}(\zeta^1,\zeta^2) \neq 0,
  \label{eq:WorldsheetMetricDoubleNull}
\end{align}
and the equation of motion \eqref{eq:NambuGotoEqGeneral} takes the form 
\begin{align}
  \pdv{x^\nu}{\zeta^2} \nabla_\nu \pdv{x^\mu}{\zeta^1}
  = 0,
  \label{eq:NambuGotoEqDoubleNull}
\end{align}
where $\nabla$ is the Levi-Civita connection on $(\Spacetime,g)$.
We note that the metric function $\gamma_{12}(\zeta^1,\zeta^2)$ does
not appear in the equation of motion.

\subsection{Cohomogeneity-one strings with a null Killing vector field}
\label{subsec:c1_string_null}

We define cohomogeneity-one strings with a null Killing vector field
and derive the equations of motion and the constraint equations.
It is convenient to use the language of differential forms,
where for a $1$-form $\TmpOneForm$, 
the exterior derivative of $\TmpOneForm$ is expressed with the Levi-Civita
connection $\nabla$ as 
\begin{align}
  \qty(\dd \TmpOneForm)_{\mu\nu}
  =
  \qty(
  \TmpOneForm_{\nu,\mu}
  -
  \TmpOneForm_{\mu, \nu}
  )
  =
  2 \nabla_{[\mu}\TmpOneForm_{\nu]},
  \label{eq:ExteriorDerivative}
\end{align}
and, for a $p$-form $B$,
the interior product with a vector field $X$ is given by
\begin{align}
  \iota_X B \coloneqq
  B(X, \underbrace{\,\cdot\, ,\dots, \,\cdot\,}_{\text{$p-1$ slots}}).
\end{align}

We assume that the spacetime $(\Spacetime, g)$ admits a null Killing vector
field $\KillingVector$, which satisfies
\begin{align}
   \nabla_{(\mu} \KillingVector_{\nu)} = 0,
   \qquad
   g_{\mu\nu} \KillingVector^{\mu} \KillingVector^{\nu} = 0.
   \label{eq:DefinitionNullKillingVector}
\end{align} 
Let $\KillingOneform$ be the metric dual $1$-form of $\KillingVector$:
\begin{align}
  \KillingOneform \coloneqq g(\KillingVector, \,\cdot\,)
  = \KillingVector_\mu \dd{x}^\mu.
  \label{eq:DefinitionKillingOneform}
\end{align}
The covariant derivative of $\KillingVector$ is given by 
\begin{align}
  \nabla_\mu \KillingVector_\nu
  =
  \frac{1}{2} \qty(\KillingTwoform)_{\mu\nu}, 
  \label{eq:CovariantDerivativeKillingOneform}
\end{align}
where Eqs.~\eqref{eq:ExteriorDerivative} and
\eqref{eq:DefinitionNullKillingVector} are used. 
It follows from Eq.~\eqref{eq:DefinitionNullKillingVector} that the null
Killing vector field $\KillingVector$ satisfies geodesic equation
\begin{align}
  \KillingVector^{\nu} \nabla_{\nu} \KillingVector^{\mu} = 0.
  \label{eq:GeodesicEq}
\end{align}
Using Eq.~\eqref{eq:CovariantDerivativeKillingOneform},
we can express this equation as 
\begin{align}
  \iota_\KillingVector \KillingTwoform = 0.
  \label{eq:GeodesicEqWithKillingOneform}
\end{align}

A cohomogeneity-one ({\COne}) string is defined as a string whose worldsheet is
tangent to a Killing vector field.
In this paper, the tangent Killing vector field is assumed to be null,
namely $\KillingVector$. 
For this tangent null Killing vector field, 
we take the null coordinate $\zeta^1$ on the worldsheet $\Worldsheet$
so that 
\begin{align}
  \pdv{x^\mu}{\zeta^1} = \KillingVector^\mu,
  \label{eq:CoordinateConditionZeta1}
\end{align}
then the equation of motion \eqref{eq:NambuGotoEqDoubleNull} is written as 
\begin{align}
  \iota_l \KillingTwoform = 0,
  \label{eq:EOMC1string}
\end{align}
where $l$ denote the other null tangent vector field $\pdv*{}{\zeta^2}$:
\begin{align}
  l^\mu \coloneqq \pdv{x^\mu}{\zeta^2}.
  \label{eq:DefinitionSymbolL}
\end{align}
In the spacetimes with $\KillingTwoform = 0$, which are known as the
{\Ppwaves} (see Sec.~\ref{sec:examples}),
the equation of motion \eqref{eq:EOMC1string} is trivial.

We now consider the case $\KillingTwoform \neq 0$.
Let $\Rank$ be $\rank \KillingTwoform$,
which is given as the maximum integer $\Rank$ such that
\begin{align}
  \qty(\KillingTwoform)^{\Rank}
  \coloneqq
  \underbrace{
  \KillingTwoform \wedge \dots \wedge \KillingTwoform
  }_{\text{$\Rank$ factors}}
  \neq
  0,
  \qquad
  2 \Rank \leqq \dim \Spacetime = \SpacetimeDimension.
  \label{eq:DefinitionRank}
\end{align}
Then it follows that 
\begin{align}
  \KillingOneform \wedge \qty(\KillingTwoform)^{\Rank} \neq 0,
  \label{eq:ConditionForKillingOneform}
\end{align}
because the equation of motion \eqref{eq:EOMC1string} implies that 
\begin{align}
  \iota_l
  \qty[\KillingOneform \wedge \qty(\KillingTwoform)^{\Rank}]
  =
  \iota_l \KillingOneform
  \,
  \qty(\KillingTwoform)^{\Rank}
  -
  \KillingOneform
  \wedge
  \qty[ \iota_{l} \qty(\KillingTwoform)^{\Rank}]
  =
  \iota_l \KillingOneform \,\,\qty(\KillingTwoform)^{\Rank}
  \neq
  0,
\end{align}
where we have used
$\iota_l \KillingOneform = g_{\mu\nu} \KillingVector^\mu l^\nu
= \gamma_{12} \neq 0$.
Eq.~\eqref{eq:ConditionForKillingOneform} and Darboux's theorem ensure the
existence of local coordinates
\begin{align}
  y^1, \dots, y^\Rank, z^1, \dots, z^\Rank,
  w, w^1, \dots, w^{\Corank - 1}
  \label{eq:DarbouxCoordinateSpacetime}
\end{align}
such that
\begin{align}
  \KillingOneform
  =
  y^1 \dd{z^1} + \dots + y^\Rank \dd{z^\Rank}
  +
  \dd{w},
  \label{eq:KillingOneformInDourbouxCoordinates}
\end{align}
where $\Corank$ is $\corank_\Spacetime \KillingTwoform$ defined by
\begin{align}
  \corank_\Spacetime \KillingTwoform
  \coloneqq
  \dim \Spacetime - 2 \rank \KillingTwoform
\end{align}
In the coordinates \eqref{eq:DarbouxCoordinateSpacetime},
the null Killing vector field $\KillingVector$ is expressed as
\begin{align}
  \KillingVector = \sum_{i = 1}^{s - 1}\KillingVector^{w^i} \pdv{}{w^i},
  \label{eq:KillingVectorInDarbox}
\end{align}
because it satisfies Eq.~\eqref{eq:GeodesicEqWithKillingOneform} and the
null condition
$g(\KillingVector,\KillingVector)
= \iota_\KillingVector \KillingOneform = 0$.
Therefore, $\Corank - 1$ must be greater than or equal to $1$, and hence 
\begin{align}
  \Corank = \corank_\Spacetime \KillingTwoform \geq 2.
  \label{eq:ConditionForCorank}
\end{align}

We consider the string worldsheet in the coordinates
\eqref{eq:DarbouxCoordinateSpacetime}.
It follows from Eqs.~\eqref{eq:CoordinateConditionZeta1} to
\eqref{eq:DefinitionSymbolL} and Eq.~\eqref{eq:KillingVectorInDarbox}
that 
\begin{align}
  \pdv{y^i}{\zeta^a}
  =
  \pdv{z^i}{\zeta^a}
  =
  0
  \quad
  (i = 1,\dots,\Rank).
\end{align}
This implies that the worldsheet $\Worldsheet$ is confined on a submanifold 
$\Submanifold$ specified by
\begin{align}
  y^i = \text{const.},
  \quad
  z^i = \text{const.}
  \quad (i = 1,\dots,\Rank).
  \label{eq:EqYZConst}
\end{align}
      %       This confinement is consistent in dimensionality
      %       because the condition \eqref{eq:ConditionForCorank} leads to 
      %       $\dim\Submanifold = \dim \Spacetime - 2 r
      %       = \corank_\Spacetime \KillingTwoform$ \geq 2.
The submanifold $\Submanifold$ is characterized by the kernels of
$\KillingTwoform_p~ (p \in \Spacetime)$,
which is defined as 
\begin{align}
  \ker \KillingTwoform_p
  \coloneqq
  \qty{
  X \in T_p(\Spacetime)
  \,|\,
  \iota_X \KillingTwoform_p = 0
  }.
\end{align}
Indeed, it follows from Eq.~\eqref{eq:KillingOneformInDourbouxCoordinates} and
\eqref{eq:EqYZConst} that, for any point $p \in \Submanifold$,
\begin{align}
  T_p(\Submanifold)
  =
  \ker \KillingTwoform_p,
  \label{eq:EqualityTangentSpaceKernel}
\end{align}
and then, we find that the submanifold $\Submanifold$ is an integral manifold
of the distribution $p \mapsto \ker \KillingTwoform_p$ for
$p \in \Spacetime$. 
In the special case that
$\corank_\Spacetime \KillingTwoform = 2$, 
which implies that $\dim \Submanifold = 2$,
the worldsheet $\Worldsheet$ is the submanifold $\Submanifold$ itself 
so that the worldsheet $\Worldsheet$ is an integral manifold of the
distribution.

We turn to the coordinate condition
\eqref{eq:CoordinateConditionZeta1} to fix the residual gauge freedom of the
worldsheet coordinate $\zeta^2$. 
Since Eq.~\eqref{eq:CoordinateConditionZeta1} shows that 
$\pdv*{}{\zeta^1}$ is a Killing vector field on the worldsheet,
the induced metric does not depend on $\zeta^1$,
and consequently, we find that the worldsheet metric is flat; indeed,
\begin{align}
  \dd{s}^2
  =
  \gamma_{ab} \dd{\zeta^a} \dd{\zeta^b}
  =
  2 \,\gamma_{12}(\zeta^2) \,\dd{\zeta^1} \dd{\zeta^2}
  =
  2 \dd{\zeta^1} \dd{\tilde{\zeta}^2},
  \quad
  \dd{\tilde{\zeta}^2}
  \coloneqq
  \gamma_{12}(\zeta^2) \dd{\zeta^2}.
  \label{eq:flat_metric_on_worldsheet}
\end{align}
This implies that we can take the worldsheet coordinate $\zeta^2$ so that  
\begin{align}
  \gamma_{12} = \iota_l \KillingOneform = 1.
  \label{eq:CoordinateConditionZeta2}
\end{align}
We impose this condition on the coordinate $\zeta^2$.
It should be noted that the coordinate $\zeta^2$ is past directed
when the coordinate $\zeta^1$, or the null Killing vector field
$\KillingVector$, is future directed.
This condition is convenient for discussing the (almost) contact structures
(see Subsec.~\ref{subsec:geometrical_structure}).

Eqs.~\eqref{eq:CoordinateConditionZeta1} and
\eqref{eq:CoordinateConditionZeta2}
and the nullness of the null tangent vector field $l$,
\begin{align}
  g_{\mu\nu} l^\mu l^\nu = 0,
  \label{eq:CoordinateConditionZeta2-2}
\end{align}
specify the worldsheet coordinates $(\zeta^1, \zeta^2)$ up to the addition of
constants. 
These are the gauge conditions to be solved
with the equation of motion \eqref{eq:EOMC1string}.

In the remainder of this paper,
the worldsheet coordinates $(\zeta^1, \zeta^2)$
are denoted by $(\lambda,\sigma)$.

\subsection{Reduction to ordinary differential equations}
\label{subsec:reduction_to_ODEs}

We construct a coordinate system in the $(\SpacetimeDimension)$-dimensional
spacetime $(\Spacetime,g)$ separate from those used in the previous
subsection, so that the equation of motion
\eqref{eq:EOMC1string} and the gauge conditions
\eqref{eq:CoordinateConditionZeta1}, 
\eqref{eq:CoordinateConditionZeta2} and
\eqref{eq:CoordinateConditionZeta2-2}
are reduced to ordinary differential equations.
The coordinate system is set up by utilizing the null Killing vector field
$\KillingVector$.
The associated one-parameter group of isometries is denoted by
$\OneParamGroupIsom$.

First, we take a hypersurface $\Hypersurface_{0}$ transversal to the orbits
of $\OneParamGroupIsom$, 
that is, each orbit intersects with $\Hypersurface_0$ once.
Then we can uniquely specify the orbits using the intersections with
$\Hypersurface_0$.
Let $\Orbit_{\Point}$ be the orbit with the intersection
$p \in \Hypersurface_0$.
The set of orbits $\qty{\Orbit_p}_{p \in \Hypersurface_0}$
fill the whole spacetime without any redundancy.

Next, we consider the hypersurfaces $\Hypersurface_v ~(v \in \mathbb{R})$
that are given by the action of $\OneParamGroupIsom$ on
$\Hypersurface_0$.
Then it follows that the set of the hypersurfaces
$\qty{\Hypersurface_v}_{v\in\mathbb{R}}$ foliates the whole spacetime.

\begin{figure}[h]
   \centering
   \includegraphics{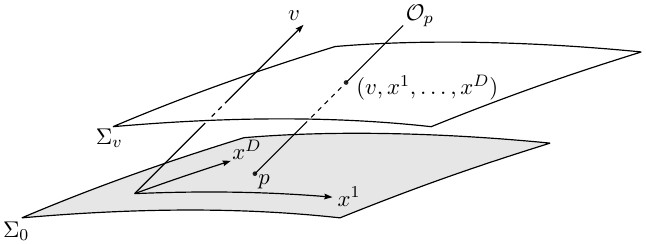}
   \caption{The coordinates $(v,x^1,\dots,x^{\HypersurfaceDimension})$
     of a point in the spacetime $\Spacetime$.
     The coordinate $v$ specifies the hypersurface $\Hypersurface_v$
     and $x^1,\dots,x^{\HypersurfaceDimension}$ specify the
     intersection $\Point$ of the orbit $\Orbit_{\Point}$
     with $\Hypersurface_0$.}
   \label{fig:coordinates}
\end{figure}

Finally, we specify each point in $\Spacetime$ by the hypersurface
$\Hypersurface_{v}$ and orbit $\Orbit_{\Point}$ on which the point lies.
Let $(x^1,\dots,x^{\HypersurfaceDimension})$ be the coordinates of the
intersection $\Point$ on $\Hypersurface_0$.
Then the point in $\Spacetime$ is labeled by
$(v,x^1,\dots,x^\HypersurfaceDimension)$ as shown in
Fig.~\ref{fig:coordinates}.
This is the coordinate system employed in this paper.
We should note that there is a freedom in choosing the hypersurface
$\Hypersurface_0$ and the coordinate system
$(x^1, \dots, x^{\HypersurfaceDimension})$ on it.

In the coordinate system $(v, x^1,\dots,x^{\HypersurfaceDimension})$,
the vector field $\pdv*{}{v}$ coincides with the null Killing vector field
$\KillingVector$ by definition. Then Eq.~\eqref{eq:CoordinateConditionZeta1}
is solved as 
\begin{align}
  v(\lambda,\sigma) = \lambda + v_0(\sigma),
  \quad
  x^{i}(\lambda,\sigma) = x^{i}(\sigma)
  \quad
  (i = 1,\dots,\HypersurfaceDimension),
  \label{eq:C1Worldsheet}
\end{align}
and the metric is written as 
\begin{align}
  \dd{s}^2
  =
  2 \Oneform_{i}(x) \dd{x}^{i} \dd{v}
  +
  \Metric_{ij}(x) \dd{x}^{i} \dd{x}^{j},
  \label{eq:SpacetimeMetricWithNullCoordinate}
\end{align}
where $\Oneform_i(x)$ and $h_{ij}(x)$ are functions of
$x^1,\dots,x^\HypersurfaceDimension$.
From these equations,
the other null tangent vector field $l \coloneqq \pdv*{}{\sigma}$ is given
by
\begin{align}
  l
  = v_0'(\sigma) \pdv{}{v} + x^i{}'(\sigma) \pdv{}{x^i}
  \eqqcolon v_0'(\sigma) \KillingVector + \hat{l}, 
  \label{eq:ComponentsOfL}
\end{align}
and the metric dual $1$-form $\KillingOneform$ of $\KillingVector$ is
given by 
\begin{align}
  \KillingOneform = \Oneform_i(x) \dd{x^i}.
  \label{eq:OneformOnHypersurface}
\end{align}
Thus the equation of motion~\eqref{eq:EOMC1string}, 
the gauge conditions~\eqref{eq:CoordinateConditionZeta2} and 
\eqref{eq:CoordinateConditionZeta2-2} reduce to
the following ordinary differential equations for $v_0(\sigma)$ and
$x^{i}(\sigma)$,
\begin{align}
   % \qty(\Oneform_{i,j} - \Oneform_{j,i}) \, x^{i}{}'
   % &=
   % 0,
\iota_{\hat{l}} \dd{\KillingOneform} = 0,
    \label{eq:reduced_eom}
  \\
  % \Oneform_{i}\, x^{i}{}'
  % &=
  % 1,
\iota_{\hat{l}} \, \KillingOneform = 1, 
    \label{eq:reduced_constraint}
  \\
  % 2 \Oneform_{i} \, x^{i}{}' v_0' + \Metric_{ij}\, x^{i}{}' x^{j}{}'
  % &=
  % 0.
2 v'_0 \, \iota_{\hat{l}} \, \KillingOneform + h(\hat{l}, \hat{l}) = 0. 
    \label{eq:reduced_constraint2}
\end{align}

\begin{figure}[h]
      \centering
      \includegraphics{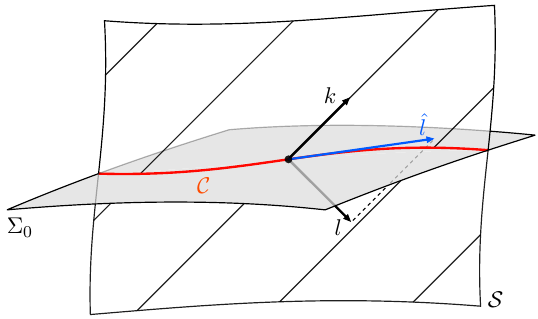}
      \caption{%
        The curve $\Curve$ given by the intersection of the worldsheet
        $\Worldsheet$ with the hypersurface $\Hypersurface_0$.
        The diagonal lines are the orbits generated by the null
        Killing vector field $\KillingVector$.
        The vector $l$ is the other null tangent vector.
        The projection $\hat{l}$, along $\KillingVector$, of $l$ onto the
        $\Hypersurface_0$ gives the tangent vector of the curve.
        }
        \label{fig:curve}
\end{figure}

We regard Eqs.~\eqref{eq:reduced_eom} and \eqref{eq:reduced_constraint}
as the equations that determine a curve
$\Curve: \sigma \mapsto (x^1(\sigma), \dots,
x^{\HypersurfaceDimension}(\sigma))$
on $\Hypersurface_0$ whose tangent vector is given by $\hat{l}$
in Eq.~\eqref{eq:ComponentsOfL}.
This curve is clearly the intersection of the worldsheet
\eqref{eq:C1Worldsheet} with the hypersurface $\Hypersurface_0$
given by $v = 0$ (see Fig.~\ref{fig:curve}).
If we have a solution $x^i(\sigma)$ of Eqs.~\eqref{eq:reduced_eom} and
\eqref{eq:reduced_constraint},
Eq.~\eqref{eq:reduced_constraint2} can be easily solved by quadrature 
\begin{align}
  v_0(\sigma)
  =
  -\frac{1}{2}
  \int
  h(\hat{l}, \hat{l})
  % \Metric_{ij} \, x^{i}{}' \, x^{j}{}'
  \dd{\sigma}.
  \label{eq:determination_v}
\end{align}
Thus our main interest is to solve Eqs.~\eqref{eq:reduced_eom}
and \eqref{eq:reduced_constraint}, 
and obtain the curve $\Curve$,
the intersection of the string worldsheet $\Worldsheet$ and the hypersurface 
$\Hypersurface_0$.

\section{General solutions and almost contact structure}
\label{sec:general_solutions}
We solve Eqs.~\eqref{eq:reduced_eom} and \eqref{eq:reduced_constraint}
on the hypersurface $\Hypersurface_0$ 
to obtain the curve $\Curve$ given by the intersection of the string
worldsheet with $\Hypersurface_0$.
The $1$-form $\KillingOneform$ appearing in these equations is
expressed only with the hypersurface coordinates
$x^1,\dots,x^\HypersurfaceDimension$ as in Eq.~\eqref{eq:OneformOnHypersurface}.
Then the $1$-form $\KillingOneform$ can be regarded as the one on the
hypersurface $\Hypersurface_0$. 
This allows us to take certain coordinates on $\Hypersurface_0$
such that Eqs.~\eqref{eq:reduced_eom} and \eqref{eq:reduced_constraint}
are easily solved.
In the case that the hypersurface $\Hypersurface_0$ is odd-dimensional,
the $1$-form $\Oneform$ provides the hypersurface $\Hypersurface_0$
with an almost contact structure, which is reviewed in Appendix
\ref{appendix:almost_contact_structure}.
With respect to the almost contact structure,
we examine the geometric structure of the general solutions.

\subsection{The case $\mathrm{d}\KillingOneform = 0$}
\label{subsec:deta_zero}
In the case $\dd{\Oneform} = 0$,
the Poincar\'{e} lemma ensures the existence of a function
$\varPhi(x^1,\dots,x^{\HypersurfaceDimension})$ on the hypersurface
$\Hypersurface_0$ such that $\Oneform = \dd{\varPhi}$.
Then, taking new local coordinates $w, w^1, \dots,
w^{\HypersurfaceDimension-1}$ on $\Hypersurface_0$ 
such that
\begin{align}
  w = \varPhi(x^1, \dots, x^{\HypersurfaceDimension}),
\end{align}
we can readily solve Eqs.~\eqref{eq:reduced_eom} and
\eqref{eq:reduced_constraint},
so that the curve
$\Curve:(w(\sigma),w^1(\sigma),\dots,w^{\HypersurfaceDimension-1}(\sigma))$
is given by
\begin{align}
  w(\sigma) = \sigma + w_{0},
  \quad
  w^{1}(\sigma),
  \dots,
  w^{\HypersurfaceDimension-1}(\sigma) : \text{arbitrary},
  \label{eq:solution_rank_zero}
\end{align}
where $w_0$ is a constant.

\subsection{The case $\mathrm{d}\KillingOneform \neq 0$}
\label{subsec:deta_non_zero}
Using the same arguments as in Subsec.~\ref{subsec:c1_string_null},
we can take local coordinates
\begin{align}
  y^1,\dots,y^\Rank, z^1,\dots,z^\Rank, w,w^1,\dots,w^{\Corank -1}
  \label{eq:DarbouxCoordinateHypersurface}
\end{align}
on the hypersurface $\Hypersurface_0$
such that
\begin{align}
  \Oneform = y^i \dd{z^i} + \dd{w},
\end{align}
where $r$ is the rank of $\dd{\Oneform}$ and $\Corank$ is the corank of
$\dd{\Oneform}$ associated with $\Hypersurface_0$,
\begin{align}
  \Corank
  = \corank_{\Hypersurface_0} \dd{\Oneform}
  = \dim \Hypersurface_0 - 2 \rank\dd{\Oneform}
  = \HypersurfaceDimension - 2 \Rank.
\end{align}
It follows from Eq.~\eqref{eq:reduced_eom} that $(2\Rank+1)$-form
$\Oneform \wedge \qty(\dd{\Oneform})^{\Rank}$ does not vanish on the
$\HypersurfaceDimension$-dimensional hypersurface $\Hypersurface_0$.
This implies that $\HypersurfaceDimension \geqq 2 \Rank + 1$,
and therefore,
\begin{align}
  \corank_{\Hypersurface_0} \dd{\Oneform} \geqq 1.
\end{align}

In the Darboux coordinates \eqref{eq:DarbouxCoordinateHypersurface},
we can readily solve
Eqs.~\eqref{eq:reduced_eom} and \eqref{eq:reduced_constraint}
so that the curve
$\Curve:(y^i(\sigma),z^i(\sigma),w(\sigma),w^{j}(\sigma))$ is given by 
\begin{align}
  y^{i}(\sigma), z^{i}(\sigma) : \text{constant}, 
\quad
  w(\sigma) = \sigma +  w_{0},
\quad
 w^{j}(\sigma):\text{arbitrary}. 
\label{eq:SolutionRankNonzero}
\end{align}
In the case that $\corank_{\Hypersurface_0} \dd{\Oneform} = 1$,
the solution does not involve any arbitrary functions.
This solution corresponds to that of the case $\corank_\Spacetime
\KillingTwoform = 2$ in Subsec.~\ref{subsec:c1_string_null},
where the worldsheet is given as an integral manifold of the distribution
$p \mapsto \ker \KillingTwoform_p ~ (p \in \Spacetime)$.

\subsection{Geometric structure of the general solutions}
\label{subsec:geometrical_structure}
In this subsection we assume that the hypersurface $\Hypersurface_0$ is
odd-dimensional. 
Then the $1$-form $\KillingOneform$ provides $\Hypersurface_0$ with an
almost contact structure $\AlmostContactStructure$:
a triplet of $(1,1)$-tensor $\Endmorphism$,
a vector field $\VectorField$ and 
the $1$-form $\Oneform$
such that 
\begin{align}
  \iota_{\VectorField} \Oneform
  &=
    1,
    \label{eq:ConditionVectorFieldAlmostContactStructure}\\
  \Endmorphism^2
  &=
    - 1 + \VectorField \otimes \Oneform.
\end{align}
Indeed, the vector field $\VectorField$ is given so that
Eq.~\eqref{eq:ConditionVectorFieldAlmostContactStructure} holds for the 
$1$-form $\Oneform$,
and the $(1,1)$-tensor $\Endmorphism$ is constructed from
$\Oneform$ and $\VectorField$ (see Appendix
\ref{appendix:almost_contact_structure}).
We should note that the vector field $\VectorField$ is not unique;
there is a freedom to add vector field $\tilde{\VectorField}$ which satisfies
$\iota_{\tilde{\VectorField}} \, \Oneform = 0$.

\begin{figure}[h]
      \centering
      \includegraphics{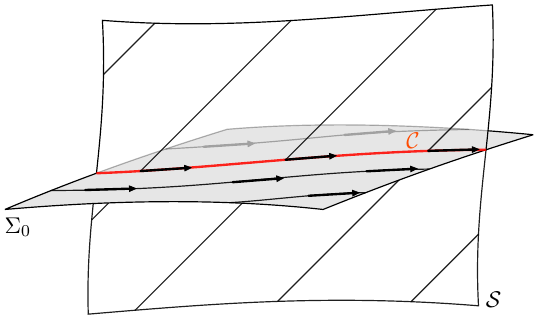}
      \caption{%
        The geometric structure of the general solutions.
        In the case that $\dim \Hypersurface_0$ is odd,
        the curve $\Curve$, the intersection of the worldsheet
        $\Worldsheet$ with the hypersurface $\Hypersurface_0$,
        is given as an integral curve of the vector field $\VectorField$
        of an almost contact structure $\AlmostContactStructure$ that satisfies
        Eqs.~\eqref{eq:ConditionVectorFieldAlmostContactStructure} and
        \eqref{eq:DefinitionReebVector}.
      }
      \label{fig:geometric_structure}
\end{figure}

We examine the general solutions obtained in the previous subsections
in terms of the almost contact structure $\AlmostContactStructure$.
To this aim, we choose the vector field $\VectorField$
so that it satisfies 
\begin{align}
  \iota_{\VectorField} \dd{\Oneform} = 0.
  \label{eq:DefinitionReebVector}
\end{align}
in addition to Eq.~\eqref{eq:ConditionVectorFieldAlmostContactStructure}.
Then it is obvious that the curve $\Curve$, the intersection of the
worldsheet $\Worldsheet$ with the hypersurface $\Hypersurface_0$,
is given as an integral curve of the vector field $\VectorField$
(see Fig.~\ref{fig:geometric_structure}).
Thus to solve the {\COne} string equations of motion
is to find an almost contact structure $\AlmostContactStructure$ such that
the vector field $\VectorField$ satisfies Eq.~\eqref{eq:DefinitionReebVector}. 
We note that the vector field $\VectorField$ is not unique
in general even though Eq.~\eqref{eq:DefinitionReebVector} is imposed.

The case $\corank_{\Hypersurface_0} \dd{\Oneform} = 1$ is special in the
sense that the almost contact structure $\AlmostContactStructure$ becomes a
contact structure. 
A $(2 \Rank + 1)$-dimensional manifold with a $1$-form $\Oneform$
of $\rank \dd{\Oneform} = \Rank$ that satisfies
$\Oneform \wedge \qty(\dd{\Oneform})^{\Rank} \neq 0$
is said to have a contact structure with a contact form $\Oneform$
(see Appendix~\ref{appendix:almost_contact_structure}). 
In this case, the vector field $\VectorField$ satisfying
Eqs.~\eqref{eq:ConditionVectorFieldAlmostContactStructure}
and \eqref{eq:DefinitionReebVector} is uniquely determined
and is called the Reeb vector field.
In the Darboux coordinates \eqref{eq:DarbouxCoordinateHypersurface}, 
the Reeb vector field $\VectorField$ is given by
\begin{align}
  \xi = \pdv{}{w}.
  \label{eq:Reeb_corank1}
\end{align}
The unique determination of the vector field $\VectorField$ corresponds to
the fact that the string solution does not include any arbitrary functions
discussed in the previous subsection.

\section{Extrinsic geometry of the worldsheet}
\label{sec:extrinsic_geometry}

We investigate the second fundamental form of the string worldsheet
$\Worldsheet$, which requires careful treatment 
because the codimension of $\Worldsheet$,
denoted by $\codim\Worldsheet$, may be equal to or larger than two.
The foundations are given in Appendix
\ref{appendix:second_fundamental_form}.

Let $N_I ~(I = 1,\dots, \Codimension)$ be independent normal vector
fields of the worldsheet $\Worldsheet$, that is, they satisfy
\begin{align}
  g(N_I, \KillingVector) = 0,
  \label{eq:NormalVectorsDefinitionK}\\
  g(N_I, l) = 0,
  \label{eq:NormalVectorsDefinitionL}
\end{align}
where $\KillingVector$ is the null Killing vector field tangent to the
worldsheet $\Worldsheet$ which is given by $\KillingVector = \pdv*{}{\lambda}$ 
and $l$ is the other null tangent vector field given by $l =
\pdv*{}{\sigma}$.
Then, the second fundamental form is characterized by
the symmetric tensors $K_I~(I=1,\dots,\Codimension)$ such that
\begin{align}
  K_{I}(X,Y)
  =
  g\qty(N_I, \nabla_X Y),
  \label{eq:ComponentsOfSecondFundamentalForm}
\end{align}
where $X$ and $Y$ are tangent vector fields on $\Worldsheet$.
This equation implies that
\begin{align}
  \qty(K_I)_{\lambda\lambda}
  \coloneqq
  K_I(\pdv{}{\lambda},\pdv{}{\lambda})
  =
  K_I(\KillingVector,\KillingVector)
  =
  g(N_I , \nabla_{\KillingVector}\KillingVector)
  =
  0,
  \label{eq:VanishingK1}
\end{align}
where we have used the fact that the null Killing vector field satisfies
$\nabla_{\KillingVector}\KillingVector = 0$.
Furthermore, it follows that
\begin{align}
  \qty(K_I)_{\lambda\sigma}
  \coloneqq
  K_I(\pdv{}{\lambda},\pdv{}{\sigma})
  =
  K_I(\KillingVector,l)
  =
  0
  \label{eq:VanishingK2}
\end{align}
because the Nambu-Goto equation of motion leads to
\begin{align}
  \Tr K_{I} = \gamma^{ab} \qty(K_I)_{ab} = 0,
\end{align}
where $\gamma^{ab}$ is the inverse of the induced metric
$\gamma_{ab}$ that has only off diagonal components
$\gamma_{\lambda\sigma} = 1$.
The only non-trivial components are
\begin{align}
  \qty(K_I)_{\sigma\sigma}
  \coloneqq
  K_I(\pdv{}{\sigma},\pdv{}{\sigma})
  =
  K_I(l,l)
  =
  g\qty(N_I, \nabla_l \,l)
  \quad
  (I = 1, \dots, \Codimension).
\end{align}
This equation implies that,
if the other null tangent vector field $l$ is geodesic,
$\qty(K_I)_{\sigma\sigma}$ also vanish.

We examine the non-trivial components $\qty(K_I)_{\sigma\sigma}$ in detail
by using the null {\COne} symmetry. 
First we take normal vector fields
$N_I~(I=1,\dots,\Codimension)$ so that $\mathcal{L}_k  N_{I} = 0$,
where $\mathcal{L}_\KillingVector$ denotes the Lie derivative along 
the null Killing vector field $\KillingVector$. 
Then, $\qty(K_I)_{\sigma\sigma}$ are determined from the values on the 
intersection with the hypersurface $\Hypersurface_0$,
namely the curve $\Curve$ on $\Hypersurface_0$
because $\qty(K_I)_{\sigma\sigma}$ are invariant
along the Killing vector field $\KillingVector$; 
\begin{align}
  \mathcal{L}_\KillingVector \qty(K_I)_{\sigma\sigma}
  =
  \mathcal{L}_\KillingVector g(N_I, \nabla_l \,l)
  =
  g(N_I, \nabla_{\mathcal{L}_\KillingVector l} \,l)
  +
  g(N_I, \nabla_l \,\mathcal{L}_\KillingVector l )
  =
  0.
\end{align}
Next, at each point $p$ on $\Curve$,
we consider two direct sum decompositions of $T_{p}\Spacetime$:
\begin{align}
  T_p \Spacetime
  =
  \Span(n) \oplus T_p \Hypersurface_0
  =
  \Span(k) \oplus T_p \Hypersurface_0,
\end{align}
where $n$ is a unit vector normal to $\Hypersurface_0$.
Let $P_n$ and $P_k$ be the projections onto $T_p \Hypersurface_0$
along $n$ and $\KillingVector$ respectively,
then, it follows that (a detailed derivation is given in
Appendix~\Ref{appendix:derivation})
\begin{align}
  \qty(K_I)_{\sigma\sigma}
  =
  g(\hat{N_I}, P_n \nabla_{\hat{l}} \, \hat{l}\,),
  \quad
  \hat{N}_I
  \coloneqq
  P_k N_I,
  \quad
  \hat{l}
  \coloneqq
  P_k l.
  \label{eq:NonTrivialComponentsKandProjection}
\end{align}
This equation can be written
by using the induced metric $h$ and its associated connection
${}^{(h)}\nabla$ on the hypersurface $\Hypersurface_0$ as follows
\begin{align}
  \qty(K_{I})_{\sigma\sigma}
  =
  h(\hat{N}_I, {}^{(h)}\nabla_{\hat{l}} \, \hat{l}\,).
  \label{eq:NonTrivialComponentsK}
\end{align}
For $\hat{N}_I$ and $\hat{l}$, 
it follows from Eq.~\eqref{eq:NormalVectorsDefinitionL} that 
\begin{align}
  h(\hat{N}_I, \hat{l}) + \frac{g(n, N_I)}{g(n,\KillingVector)} = 0.
  \label{eq:Orthogonality}
\end{align}
Here we note that $\hat{l}$ is the tangent vector to the curve
$\Curve$ as depicted in Fig.~\ref{fig:curve}.
Then, from Eqs.~\eqref{eq:NonTrivialComponentsK} and
\eqref{eq:Orthogonality},
we find two special cases where we can discuss the non-trivial
components $\qty(K_I)_{\sigma\sigma}$ in relation to the geometry of the 
hypersurface $\Hypersurface_0$.

The first case is when the hypersurface $\Hypersurface_0$ is orthogonal
to the worldsheet $\Worldsheet$, namely $g(n,N_I) = 0$.
In this case, it follows form Eq.~\eqref{eq:Orthogonality} that 
\begin{align}
  h(\hat{N}_I, \hat{l}\,) = 0.
  \label{eq:OrthogonalityOfNandC}
\end{align}
Using this equation and Eq.~\eqref{eq:NonTrivialComponentsKandProjection}, 
we find that $\qty(K_I)_{\sigma\sigma}$ vanish if and only 
if the curve $\Curve$ is a geodesic on $\Hypersurface_0$, 
namely $\hat{l}$ satisfies 
\begin{align}
  {}^{(h)}\nabla_{\hat{l}} \, \hat{l} 
  =
  \varphi \, \hat{l}
  \label{eq:GeodesicEqForC}
\end{align}
for some function $\varphi$.

The second case is when $\corank_{\Hypersurface_0} \dd{\Oneform} = 1$,
where the almost contact structure $\AlmostContactStructure$ on 
$\Hypersurface_0$ becomes the contact structure
and the curve $\Curve$ is given as an integral curve of the Reeb vector
field $\VectorField$.
In this case, we find that the non-trivial components
$\qty(K_I)_{\sigma\sigma}$ vanish if the Reeb vector field $\VectorField$ is
a Killing vector field with a constant norm. 
Indeed, a Killing vector field with a constant norm always satisfies the
geodesic equation, and thus the Reeb vector field $\VectorField$ satisfies 
\begin{align}
  {}^{(h)}\nabla_\VectorField \, \VectorField = 0.
  \label{eq:GeodesicEqForReeb}
\end{align}
This implies that the curve $\Curve$ satisfies the
geodesic equation \eqref{eq:GeodesicEqForC} with
$\varphi = 0$ and hence $\qty(K_I)_{\sigma\sigma}$ vanish.

\section{Examples}
\label{sec:examples}
In this section we apply the methods described
in Secs.~\ref{sec:equations_of_motion} and \ref{sec:general_solutions}
in three four-dimensional spacetimes that admit a null Killing vector field
$\KillingVector$. 
The first spacetime is the
\emph{plane-fronted gravitational waves with parallel rays} ({\Ppwaves}).
The {\Ppwaves} is defined as a spacetime with a null vector field 
$\KillingVector$ that satisfies $\nabla_{\nu}\KillingVector_{\mu} = 0$, 
and thus admits a null Killing vector field.
The condition $\nabla_{\nu}\KillingVector_{\mu} = 0$ implies that the
metric dual $1$-form of $\KillingVector$, namely $\Oneform$, satisfies
$\dd{\Oneform} = 0$.
The second and third spacetimes are the Einstein static universe and the
G\"{o}del universe.
Both spacetimes are homogeneous in space and time.
Furthermore, they admit spacelike and timelike Killing vector fields of
constant norm.
Thus they admit null Killing vector fields.
It will be shown that, in both spacetimes,
the metric dual $1$-form $\Oneform$ satisfies
$\dd{\Oneform} \neq 0$ and $\rank(\dd{\Oneform}) = 1$.

\subsection{The {\Ppwaves}}
\label{subsec:pp_waves}
The metric of the {\Ppwaves} is written in the Brinkmann coordinates as
\begin{align}
  \dd{s}^2
  =
  2 \dd{w} \dd{v}
  +
  2 H(w,w^i) \qty(\dd{w})^2
  +
  \qty(\dd{w^1})^2
  +
  \qty(\dd{w^2})^2,
  \label{eq:MetricPpwaves}
\end{align}
where $H(w,w^i)$ is a function of $w,w^1,w^2$ determined by the
Einstein equations \cite{stephani2009exact}.
This metric form shows that the Brinkmann coordinates $(v, w, w^1, w^2)$
are suitable for applying the methods of
Subsec.~\ref{subsec:reduction_to_ODEs} and \ref{subsec:deta_zero}.
Indeed, the null Killing vector field $\KillingVector$,
the hypersurface $\Hypersurface_0$ and the metric dual $1$-form of
$\KillingVector$ are given by $\pdv*{}{v}$, $v = 0$ and
$\KillingOneform = \dd{w}$ respectively.
Thus, Eqs.~\eqref{eq:C1Worldsheet} and \eqref{eq:solution_rank_zero}
can be used, and the worldsheet is obtained as follows
\begin{align}
  v(\lambda,\sigma) = \lambda + v_0(\sigma),
  \quad
  w(\lambda,\sigma) = \sigma + w_0,
  \quad
  w^i(\lambda,\sigma) = w^i(\sigma)
  \quad
  (i = 1,2),
  \label{eq:WorldsheetPp}
\end{align}
where $w^i(\sigma)$ are arbitrary functions and
$v_0(\sigma)$ is determined by Eq.~\eqref{eq:determination_v}.

We examine the second fundamental form of the worldsheet.
Since the codimension of the worldsheet is two,
there are two independent normal vector fields $N_1, N_2$, 
which are for example given by
\begin{align}
  N_{I}
  =
  - w^{I}{}'(\sigma) \, \pdv{}{v} + \pdv{}{w^I}
  \quad
  (I = 1,2).
  \label{eq:NormalVectorsPpwaves}
\end{align}
For these normal vector fields,  
the non-trivial components of the second fundamental form
given by Eq.~\eqref{eq:NonTrivialComponentsK} are
\begin{align}
  (K_I)_{\sigma\sigma}
  =
  w^I{}'' - \pdv{H}{w^I}
  \quad
  (I = 1,2).
\end{align}

We also examine the twist potential $\TwistPotential_I{}^{J}$,
which are the $1$-forms on the worldsheet $\Worldsheet$ defined by
Eq.~\eqref{eq:definition_twist_potential}.
The twist potential requires that two normal vector fields are orthonormal.
The normal vector fields $N_1, N_2$ given by
Eq.~\eqref{eq:NormalVectorsPpwaves} satisfy the requirement.
Using the formula \eqref{eq:twist_potential_formula} 
we obtain 
\begin{align}
  \TwistPotential_I{}^J  =  0 \quad (I,J = 1,2).
\end{align}
Then, from Eq.~\eqref{eq:definition_twist_potential}, we have
\begin{align}
  g(\nabla_\KillingVector N_1, N_2) = 0,
  \quad
  g(\nabla_l N_1, N_2) = 0.
\end{align}
These equations imply that the worldsheet does not twist in the sense that
the normal vector fields do not rotate when they are parallelly transported
along the null directions $\KillingVector, l$ on the worldsheet.

\begin{figure}[h]
      \centering
      \includegraphics[]{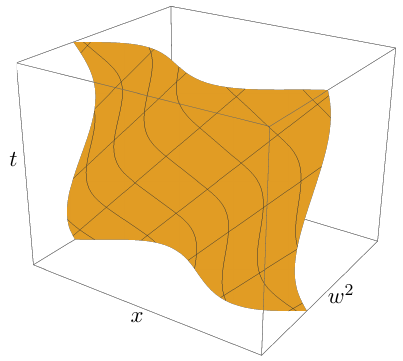}
      \caption{
        The string worldsheet specified by
        Eq.~\eqref{eq:SpecificationOfSolution}.
        The worldsheet is in effect embedded in three-dimensional
          spacetime $(w^1 = 0)$.
        }
      \label{fig:world_sheet_minkowski}
\end{figure}

For more intuitive understanding of the extrinsic geometry, 
let us consider a specific case
\begin{align}
  H(w,w^i) = 0,
  \quad
  w^1(\sigma) = 0,
  \quad
  w^2(\sigma) = \sin \sigma.
  \label{eq:SpecificationOfSolution}
\end{align}
The first condition $H(w,w^i) = 0 $ implies that the spacetime is flat; 
in fact, the metric \eqref{eq:MetricPpwaves} becomes
\begin{align}
  \dd{s}^2
  =
  -\dd{t}^2 + \dd{x}^2 + \qty(\dd{w^1})^2 + \qty(\dd{w^2})^2,
  \quad
  t \coloneq \frac{1}{\sqrt{2}} (v - w),
  \quad
  x \coloneq \frac{1}{\sqrt{2}} (v + w).
\end{align}
The second one means that the worldsheet is confined on the hyperplane
$w^{1}=0$. 
Therefore the worldsheet can be depicted in the $3$-dimensional flat spacetime
as in Fig.~\ref{fig:world_sheet_minkowski}.

The straight lines in Fig.~\ref{fig:world_sheet_minkowski},
which are geodesics,
are the orbits of the null Killing vector field
$\KillingVector = \pdv*{}{\lambda}$. 
The curved lines are the orbits of the other tangent null vector field
$l = \pdv*{}{\sigma}$.
The worldsheet is curved along $l$, but not along $\KillingVector$.
This is consistent with the second fundamental form:
$\qty(K_1)_{\sigma\sigma} = \sin\sigma \neq 0,
\qty(K_1)_{\lambda\lambda} = 0.$

\subsection{The Einstein static universe}
\label{subsec:einstein_static_universe}
The Einstein static universe is a closed
Friedmann-Lema\^{i}tre-Robertson-Walker universe with a constant scale
factor $a (>0)$. 
The metric is given by 
\begin{align}
  \dd{s}^2
  =
  a^2
  \qty[
  - \dd{t}^2
  +
  \dd{\theta}^2
  +
  \sin^2\theta \dd{\phi}^2
  +
  \qty(
  \dd{\psi} - \cos \theta \dd{\phi}
  )^2
  ],
  \label{eq:MetricESU}
\end{align}
where the spatial coordinates are chosen so that they reflect the Hopf fibration 
of $S^3$.
It is clear that $\pdv*{}{t} \pm \pdv*{}{\psi}$
or their constant multiples are null Killing vector fields.
In order to clarify the influence of having two independent null Killing vector
fields, we assume, for a while, that the scale factor $a$ is a function of 
$\psi - t$.
Then the metric \eqref{eq:MetricESU} admits only one  null Killing vector
field $\KillingVector$ of the form
\begin{align}
  \KillingVector = c \qty(\pdv{}{t} + \pdv{}{\psi}),
  \label{eq:NullKillingVectorFieldESU}
\end{align}
where $c (\neq 0)$ is a constant.

For the null Killing vector field \eqref{eq:NullKillingVectorFieldESU},
the metric dual $1$-form
\eqref{eq:DefinitionKillingOneform} is 
\begin{align}
  \KillingOneform
  =
  c a^2 \qty(-\dd{t} + \dd{\psi} - \cos\theta \dd{\phi}).
  \label{eq:KillingOneformInESU}
\end{align}
We readily find that
\begin{align}
  \rank \KillingTwoform = 1,
  \quad
  \corank_\Spacetime  \KillingTwoform = 2,
  \quad
  \KillingOneform \wedge \qty(\KillingTwoform)^1 \neq 0.
  \label{eq:PropertiesOfOneformESU}
\end{align}
Then, as discussed in Subsec.~\ref{subsec:c1_string_null},
the worldsheet is given as an integral manifold of the distribution 
$p \mapsto \ker \KillingTwoform_p ~(p \in \Spacetime)$.
The following two vector fields give a basis of the kernel
$\ker \KillingTwoform_p$ at each point $p$:
\begin{align}
  \pdv{}{t} -  \frac{2a'\cot\theta}{a} \pdv{}{\theta},
  \quad
  \pdv{}{\psi} + \frac{2a'\cot\theta}{a}  \pdv{}{\theta},
\end{align}
where $a'$ is the derivative of the scale factor.
In the case $a' = 0$, namely the case of the Einstein static universe,
these vector fields are just $\pdv*{}{t}$ and $\pdv*{}{\psi}$, and then the
worldsheet may be simply specified by $\theta, \phi = \text{const.}$

Let us obtain the worldsheet for the case $a' \neq 0$
by applying the methods
described in Subsecs.~\ref{subsec:reduction_to_ODEs}
and \ref{subsec:deta_non_zero}.
First we take the hypersurface $t = 0$ as $\Hypersurface_0$, 
which is transversal to the null Killing vector field $\KillingVector$.
Next we take coordinates $x^1,x^2,x^3$ on $\Hypersurface_0$
so that the spacetime coordinates of a point on $\Hypersurface_0$ are 
given by $(t,\theta, \phi, \psi) = (0, x^1, x^2, x^3)$.
Then the action of the $1$-parameter group of isometries
$\OneParamGroupIsom$ is given by
\begin{align}
  \OneParamGroupIsom:
  (0, x^1, x^2, x^3)
  \mapsto
  (c v, x^1, x^2, x^3 + c v).
\end{align}
Let $(t,\theta, \phi, \psi)$ be the coordinates of
the point $(c v, x^1, x^2, x^3 + c v)$, 
then the coordinate transformation is given by
\begin{align}
  t = c v,
  \quad
  \theta = x^1,
  \quad
  \phi = x^2,
  \quad
  \psi = x^3 + c v.
\end{align}
In these coordinates, the metric is written as
\begin{align}
  \dd{s}^2
  =
  2 c a^2 \qty(\dd{x^3} - \cos x^1 \dd{x^2}) \dd{v}
  +
  a^2
  \qty[
  \qty(\dd{x^1})^2
  +
  \sin^2 x^1 \qty(\dd{x}^2)^2
  +
  \qty(
  \dd{x^3} - \cos x^1 \dd{x^2}
  )^2
  ],
\end{align}
where we note that the scale factor $a$ becomes a function of
$x^3 (= \psi - t)$. 
The metric dual $1$-form \eqref{eq:KillingOneformInESU} is given by  
\begin{align}
  \Oneform
  =
  c a^2\qty(\dd{x^3} - \cos x^1 \dd{x^2}).
  \label{eq:contact_form_ESU}
\end{align}
This $1$-form is regarded as the one furnished on the hypersurface
$\Hypersurface_0$ and satisfies $\Oneform \wedge \dd{\Oneform} \neq 0$.
Therefore, Darboux's theorem ensures that the hypersurface $\Hypersurface_0$
admits local coordinates $y,z,w$ such that 
\begin{align}
  \Oneform = y \dd{z} + \dd{w}.
  \label{eq:contact_form_ESU_darboux}
\end{align}
The coordinate transformation is, for example, given by
\begin{align}
  y = - c a^2 \cos x^1,
  \quad
  z = x^2,
  \quad
  w = \int c a^2(x^3) \dd{x^3} \eqqcolon f(x^3).
  \label{eq:darboux_coordinate_ESU}
\end{align}
In these coordinates,
the induced metric $h$ on $\Hypersurface_0$ is written as
\begin{align}
      \dd{s}_{\Hypersurface_0}^2
      =
      \frac{\tilde{a}^2}{c^2 \tilde{a}^4 - y^2}
      \qty(\dd{y} - 2y\frac{\tilde{a}'}{\tilde{a}}\dd{w})^2
      +
      \frac{c^2\tilde{a}^4 - y^2}{c^2 \tilde{a}^2} \dd{z}^2
      +
      \frac{1}{c^2\tilde{a}^2} \qty(y \dd{z} + \dd{w})^2,
\end{align}
where $\tilde{a}$ is the function of $w$ such that
$\tilde{a}(w)= a(f^{-1}(w))$,
and the worldsheet is given by Eqs.~\eqref{eq:C1Worldsheet}
and \eqref{eq:SolutionRankNonzero} as
\begin{align}
  v(\lambda, \sigma) = \lambda + v_0(\sigma),
  \quad
  y(\lambda, \sigma) = y_0,
  \quad
  z(\lambda, \sigma) = z_0,
  \quad
  w(\lambda, \sigma) = \sigma + w_0,
  \label{eq:C1WorldsheetESU}
\end{align}
where $y_0, z_0, w_0$ are constants and,
from Eq.~\eqref{eq:determination_v}, 
$v_0(\sigma)$ is determined as
\begin{align}
  v_0(\sigma)
  =
  -
  \frac{1}{2}
  \int h_{ww} \dd{\sigma}
  =
  -
  \frac{1}{2}
  \int
  \qty(
  \frac{4 \, y^2 \, \tilde{a}'{}^2}{c^2 \tilde{a}^4 - y^2}
  +
  \frac{1}{c^2 \tilde{a}^2}
  )
  \dd{\sigma}.
  \label{eq:DeterminationVESU}
\end{align}

We now examine the second fundamental form of the worldsheet
\eqref{eq:C1WorldsheetESU}.
Since the worldsheet is simply given by $y = y_0$ and $z = z_0$,
we take two normal vector fields $N_1$ and $N_2$ so that their
metric dual $1$-forms are $\dd{y}$ and $\dd{z}$.
For these $N_1, N_2$, the non-trivial components of the second fundamental
form are given by Eq.~\eqref{eq:NonTrivialComponentsKandProjection} as
\begin{align}
  \qty(K_1)_{\sigma\sigma}
  &=
    -\frac{2y_0}{\tilde{a}^2\qty(c^2 \tilde{a}^4 - y_0^2)^2}
    \qty[
    -(c^2 \tilde{a}^4 - y_0^2)(c^2 \tilde{a}^4 + 2y_0^2)\, \tilde{a}'{}^2
    +
    4 c^2 \,  y_0^4 \, \tilde{a}^2 \, \tilde{a}'{}^4
    +
    (c^2 \tilde{a}^4 - y_0^2) \, \tilde{a} \, \tilde{a}''
    ],
  \\
  \qty(K_2)_{\sigma\sigma}
  &=
    -
    \frac{y_0\tilde{a}'}{\tilde{a} \qty(c^2\, \tilde{a}^4 - y_0^2)^2}
    \qty[y_0^2 \qty(4c^2 \, \tilde{a}^2 \, \tilde{a}'{}^2 + 1) - c^2 \,\tilde{a}^4],
\end{align}
where the function $\tilde{a}$ is evaluated on the worldsheet
\eqref{eq:C1WorldsheetESU}, that is,
$\tilde{a} = \tilde{a}(\sigma + w^0)$.
From these expressions,
we readily find that if the scale factor $a$ is constant, which is the
case of the Einstein static universe, the second fundamental form
vanishes.
Conversely, it is easily shown that if the second fundamental form of
\emph{every} worldsheet vanishes, the scale factor has to be constant.
Therefore, the Einstein static universe is the only
spacetime with the metric \eqref{eq:MetricESU}
that permits \emph{every} worldsheet of null {\COne} symmetry to
have a vanishing second fundamental form.
The reason for the vanishing of the non-trivial components in the
Einstein static universe is that the null vector field $l =
\pdv*{}{\sigma}$ tangent to the worldsheet agrees with a constant multiple 
of the other null Killing vector field $\pdv*{t} - \pdv*{\psi}$:
in fact, from Eqs.~\eqref{eq:C1WorldsheetESU} and
\eqref{eq:DeterminationVESU}, 
\begin{align}
  l = -\frac{1}{2c\,a^2} \qty(\pdv{}{t} - \pdv{}{\psi}).
\end{align}
In the remainder of this subsection,
we only consider the case $a = \text{const.}$, namely the case of the
Einstein static universe, 
where the induced metric $h$ is given by
\begin{align}
  \dd{s}_{\Hypersurface_0}^2
  =
  \frac{a^2}{c^2 a^4 - y^2}
  \dd{y}^2
  +
  \frac{c^2 a^4 - y^2}{c^2 a^2} \dd{z}^2
  +
  \frac{1}{c^2 a^2} \qty(y \dd{z} + \dd{w})^2,
  \label{eq:InducedMetricESU}
\end{align}
and the normal vector field $N_1, N_2$ used above are given by
\begin{align}
  N_1 = \frac{c^2a^4-y^2}{a^2} \pdv{}{y},
  \quad
  N_2 = \frac{c^2a^2}{c^2 a^4 - y^2}\qty(\pdv{}{z} - y \pdv{}{w}).
  \label{eq:NormalVectorsESU}
\end{align}

We discuss the vanishing of the second fundamental form described above
from two perspectives. 
The first is the orthogonality of $\Hypersurface_0$ and the
worldsheet $\Worldsheet$.
It follows from Eq.~\eqref{eq:NormalVectorsESU}
that $\Hypersurface_0$ is orthogonal to $\Worldsheet$,
namely $g(n, N_I) = 0$, where $n$ is a unit normal vector field of
$\Hypersurface_0$ given by $v = 0$ in $(v,y,z,w)$ coordinates.
Therefore, as discussed in Sec.~\ref{sec:extrinsic_geometry},
the vanishing of the second fundamental form
implies that the curve $\Curve$, which is the section of 
$\Worldsheet$ with $\Hypersurface_0$, is a geodesic on
$\Hypersurface_0$.
Indeed, from Eq.~\eqref{eq:C1WorldsheetESU},
$\Curve$ on $\Hypersurface_0$ is given by
$(y(\sigma), z(\sigma), w(\sigma))  = (y_0,  z_0, \sigma + w_0)$,
and the tangent vector $\pdv*{}{w}$ is geodesic for the induced metric
\eqref{eq:InducedMetricESU} because $\pdv*{}{w}$ is a Killing vector field of
a constant norm.
The second is a contact structure $\AlmostContactStructure$
on $\Hypersurface_0$.
It follows from Eq.~\eqref{eq:PropertiesOfOneformESU} that 
\begin{align}
  \corank_{\Hypersurface_0} \dd{\Oneform} = 1.
\end{align}
Thus, the hypersurface $\Hypersurface_0$ has a contact
structure $\AlmostContactStructure$.
As discussed in Sec.~\ref{sec:extrinsic_geometry},
a sufficient condition for the second fundamental form to vanish is
that the Reeb vector field $\VectorField$ satisfies the geodesic equation 
\eqref{eq:GeodesicEqForReeb},
i.e., ${}^{(h)}\nabla_\VectorField \, \VectorField = 0$.
This condition is actually satisfied.
In fact, in the Darboux coordinates $(y,z,w)$,
the Reeb vector field $\VectorField$ is given by
Eq.~\eqref{eq:Reeb_corank1}, i.e., $\VectorField = \pdv*{}{w}$
and then, is clearly a Killing vector field of a constant norm
with respect to the induced metric \eqref{eq:InducedMetricESU}.
This implies that  $\VectorField$ satisfies the geodesic equation.

We note that the induced metric $\Metric$ given by
\eqref{eq:InducedMetricESU} is not generally compatible with the contact
structure $\AlmostContactStructure$ on $\Hypersurface_0$. 
However, if we set the constant $c$ of the null Killing vector field
\eqref{eq:NullKillingVectorFieldESU} to be
$\pm 1/a$, the induced metric becomes compatible to
$\AlmostContactStructure$, 
that is, $\Metric$ satisfies
\begin{align}
  \Metric(\Endmorphism V_1, \Endmorphism V_2)
  =
  \Metric(V_1, V_2) - \Oneform(V_1) \, \Oneform(V_2)
  \label{eq:Compatibility}
\end{align}
for arbitrary vector fields $V_1, V_2$.
Indeed, if we take an orthonormal basis
$\qty{\VectorField_1,\VectorField_2,\VectorField_3}$
for the induced metric \eqref{eq:InducedMetricESU} as 
\begin{align}
  \VectorField_1
  =
  \frac{\sqrt{a^2 - y^2}}{a} \pdv{}{y},
  \quad
  \VectorField_2
  =
  \frac{1}{\sqrt{a^2 - y^2}}
  \qty(\pdv{}{z} - y \pdv{}{w}),
  \quad
  \VectorField_3
  = \VectorField = \pdv{w},
  \label{eq:OrthonormalBasisESU}
\end{align}
such that $\Oneform(\VectorField_1) = \Oneform(\VectorField_2) = 0$,
and define the $(1,1)$-tensor $\Endmorphism$
\begin{align}
  \Endmorphism^I{}_J
  \coloneqq
  \Endmorphism(\Oneform^I, \VectorField_J)
  =
  \mqty(
  0 & -1 & 0 \\
  1 & 0 & 0 \\
  0 & 0 & 0),
\end{align}
where $\Oneform^I~(I=1,2,3)$ are the dual $1$-forms
to $\VectorField_I$,
then it is readily verified that the induced metric $h$
satisfies the compatibility condition \eqref{eq:Compatibility}.

We next examine the twist potential $\TwistPotential_I{}^{J}$
of the worldsheet by assuming $c = 1/a$.
For this aim, we have to take normal vector fields $N_1, N_2$
so that they are orthonormal, namely $g(N_I, N_J) = \delta_{IJ}$.
This requirement is satisfied by taking $\VectorField_1, \VectorField_2$
of Eq.~\eqref{eq:OrthonormalBasisESU} to $N_1, N_2$.
These vector fields satisfy
$\mathcal{L}_{\KillingVector} N_I = \mathcal{L}_l N_I = 0$. 
For these normal vector fields,
Eq.~\eqref{eq:twist_potential_formula} reads
\begin{align}
  \qty(\TwistPotential_I{}^J)_{\lambda}
  =
  \TwistPotential_I{}^J(\pdv{}{\lambda})
  =
  \frac{1}{2a}
  \mqty(
  0 & 1 \\
  -1 & 0
  ),
       \quad
       \qty(\TwistPotential_I{}^J)_{\sigma}
       =
       \TwistPotential_I{}^J(\pdv{}{\sigma})
       =    
       \frac{1}{4a}
       \mqty(%
       0 & 1 \\
  -1 & 0
       ),
\end{align}
and then, from Eq.~\eqref{eq:definition_twist_potential}, we have
\begin{align}
  g(\nabla_{\KillingVector}N_1, N_2) = \frac{1}{2a},
  \quad
  g(\nabla_{l}N_1, N_2) = \frac{1}{4a}.
\end{align}
These equations imply that the worldsheet twists in the sense that the
normal vector fields rotate when they are parallelly propagated along the
null directions while the Lie derivatives vanish.
We also find that for the unit timelike and spacelike vector fields
tangent to the worldsheet $\Worldsheet$ 
\begin{align}
  e_0
  \coloneqq
  \frac{1}{a} \pdv{}{t}
  =
  \frac{1}{2a}\KillingVector - al,
  \quad
  e_1
  \coloneqq
  \frac{1}{a} \pdv{}{\psi}
  =
  \frac{1}{2a}\KillingVector + al,
\end{align}
which are orthogonal  to each other,
it holds that 
\begin{align}
  g(\nabla_{e_0}N_1, N_2) = 0,
  \quad
  g(\nabla_{e_1}N_1, N_2) = \frac{1}{2a}.
\end{align}
This implies the twist of the worldsheet comes from the $e_1$ direction,
which is the direction of the $S^1$ fibers in the Hopf fibration of $S^3$.
The value $1/(2a)$ is just the half of the Hodge dual of the $3$-form $\Oneform
\wedge \dd{\Oneform}$ in $\Hypersurface_0$.

\subsection{The G\"{o}del universe}
\label{subsec:godel}
We start with the following metric
\begin{align}
  \dd{s}^2
  =
  a^2
  \qty[
  -
  \qty(
  \dd{T} + e^{Y} \dd{Z}
  )^2
  +
  \dd{Y}^2
  +
  \frac{1}{2} e^{2Y} \dd{Z}^2
  +
  \dd{W}^2
  ],
  \label{eq:MetricGodel}
\end{align}
where $a$ is a function of $T - W$.
This metric admits a null Killing vector field $\KillingVector$ of the form
\begin{align}
  \KillingVector
  =
  c \qty(\pdv{}{T} + \pdv{}{W}),
\end{align}
where $c$ is a constant.
In the special case that $a$ is constant,
the metric describes the G\"{o}del universe
and also admits another null Killing vector field
given by the constant multiple of $\pdv*{}{T} - \pdv*{}{W}$.

The metric dual $1$-form of the null Killing vector field $\KillingVector$
is given by 
\begin{align}
  \KillingOneform
  =
  c a^2 \qty(- \dd{T} - e^Y \dd{Z} + \dd{W}).
\end{align}
Using the same arguments as in the previous subsection,
we find that the string worldsheet is given as an integral manifold
of $\ker \KillingTwoform$, 
which is tangent to the vector fields
\begin{align}
  \pdv{}{T} - 2\frac{a'}{a} \pdv{}{Y},
  \quad
  \pdv{}{W} + 2\frac{a'}{a} \pdv{}{Y}.
\end{align}
In the case that $a' = 0$, namely the case of the G\"{o}del universe,
these vector fields become $\pdv*{}{T}$ and $\pdv*{}{W}$,
and then, the worldsheet can simply be given by $Y, Z = \text{const.}$

The worldsheet in the case $a' \neq 0$ is also exactly obtained in the same
way as in the previous subsection. 
The hypersurface $\Hypersurface_0$ is taken as
$W = 0$ and the spacetime coordinates $(v,x^1,x^2,x^3)$
are taken so that
\begin{align}
  T = x^1 + cv,
  \quad
  Y = x^2,
  \quad
  Z = x^3,
  \quad  
  W = cv.
\end{align}
In these coordinates, the metric dual $1$-form $\Oneform$ is
\begin{align}
  \Oneform
  =
  - c a^2(x^1) \qty(\dd{x^1} + e^{x^2} \dd{x^3}).
\end{align}
It is readily found that the hypersurface $\Hypersurface_0$ admits the
Darboux coordinates $y, z, w$ such that $\Oneform = y\dd{z} + \dd{w}$.
The coordinate transformation is, for example, given by
\begin{align}
  y = -c a^2(x^1) e^{x^2},
  \quad
  z = x^3,
  \quad
  w = -c \int a^2(x^1) \dd{x^1} \eqqcolon F(x^1).
\end{align}
Then, from Eqs.~\eqref{eq:C1Worldsheet} and \eqref{eq:SolutionRankNonzero},
the worldsheet is given by
\begin{align}
  v(\lambda, \sigma) = \lambda  + v_0(\sigma),
  \quad
  y(\lambda, \sigma) = y_0,
  \quad
  z(\lambda, \sigma) = z_0,
  \quad
  w(\lambda, \sigma) = \sigma + w_0,
\end{align}
where $y_0, z_0, w_0$ are constants and
$v_0(\sigma)$ is a function determined by Eq.~\eqref{eq:determination_v}.

The non-trivial components of the second fundamental form 
are computed as
\begin{align}
  \qty(K_1)_{\sigma\sigma}
  =
  -\frac{2 y \qty[3\qty(\tilde{a}')^2 + 8c^2 \tilde{a}^2 \qty(\tilde{a}')^4
  + \tilde{a} \tilde{a}'']}{\tilde{a}^2},
  \quad
  \qty(K_2)_{\sigma\sigma}
  =
  \frac{2\tilde{a}' \qty[1 + 4c^2 \tilde{a}^2 \qty(\tilde{a}')^4]}{y \tilde{a}},
\end{align}
where $\tilde{a}$ is a function of $w$ such that
$\tilde{a}(w) = a(F^{-1}(w))$.
From this equation, we find that the second fundamental form  vanishes if
and only if $\tilde{a}$ is constant.
This implies that the G\"{o}del universe is the only spacetime with the
metric \eqref{eq:MetricGodel} in which every string with a null symmetry has
a vanishing second fundamental form.
In the remainder of this subsection, we only consider the case that $a$ is
constant, namely the case of the G\"{o}del universe.

As we have examined so far,
the string worldsheet in the G\"{o}del universe can be discussed in the
same way as in the Einstein static universe.
The vanishing of the second fundamental form in the G\"{o}del
universe can also be discussed in the same way as in the previous
subsection. 
Therefore, we will only mention the differences.
The first is the squared norm of the Reeb vector field $\VectorField$,
or equivalently the contact form $\Oneform$.
In the G\"{o}del universe a timelike hypersurface given by $W = 0$ is
taken as $\Hypersurface_0$ and the Reeb vector field $\VectorField$ is
timelike,
while in the Einstein static universe a spacelike hypersurface given by
$t=0$ is taken and $\VectorField$ is spacelike. 
Therefore, we have to use the other sign of $\epsilon$ for the
compatibility condition \eqref{eq:condition_almost_contact_metric_Lorentz}
in the G\"{o}del universe.
The second is the direction of the twist of the worldsheet $\Worldsheet$.
The twist potential $\TwistPotential_I^J$ is computed as
\begin{align}
  \qty(\TwistPotential_I{}^J)_{\lambda}
  =
  \frac{1}{\sqrt{2}a}
  \mqty(%
   0 & 1 \\
  -1 & 0
       ),
       \quad
       \qty(\TwistPotential_I{}^J)_{\sigma}
       =    
       - \frac{1}{2\sqrt{2}a}
       \mqty(%
       0 & 1 \\
  -1 & 0
       ),
\end{align}
where the normal vector fields $N_1, N_2$ that satisfies $g(N_I,N_J) = \delta_{IJ}$
are taken as
\begin{align}
  N_1
  =
  \frac{y}{a} \pdv{}{y},
  \quad
  N_2
  =
  \frac{\sqrt{2}}{y}
  \qty(\pdv{}{z} - y \pdv{}{w}).
  \label{eq:normal_vectors_Godel}
\end{align}
Then, it follows that 
\begin{align}
  g(\nabla_{e_0}N_1, N_2) = \frac{1}{\sqrt{2}a},
  \quad
  g(\nabla_{e_1}N_1, N_2) = 0.
\end{align}
for the unit timelike and spacelike vector fields
tangent to $\Worldsheet$:
\begin{align}
  e_0
  \coloneqq
  \frac{1}{a} \pdv{}{T}
  =
  \frac{1}{2}\pdv{}{\lambda} - \pdv{}{\sigma},
  \quad
  e_1
  \coloneqq
  \frac{1}{a} \pdv{}{W}
  =
  \frac{1}{2}\pdv{}{\lambda} + \pdv{}{\sigma}.
\end{align}
This result implies that the worldsheet twists along the timelike
direction $\pdv*{}{T}$ in the G\"{o}del universe
while it twists along the spacelike direction $\pdv*{}{\psi}$
in the Einstein static universe. 
The value $1/(\sqrt{2} a)$ is also just the half of 
the Hodge dual of the $3$-form $\Oneform \wedge \dd{\Oneform}$ in
$\Hypersurface_0$.
This result is the same as in the case of the Einstein static universe.

\section{Conclusion}
\label{sec:conclusion}
We have investigated the dynamics of the Nambu-Goto strings with a null
symmetry in curved spacetimes $\Spacetime$ that admit a null Killing vector field
$\KillingVector$.  
The null symmetry, or null cohomogeneity one ({\COne}) symmetry, means that
the null Killing vector field $\KillingVector$ is tangent to the string worldsheet.
The equation of motion and the gauge conditions are given in terms of the
metric dual $1$-form
$\KillingOneform \coloneqq g_{\mu\nu}\KillingVector^\mu \dd{x^\nu} $.
In the special case $\corank_{\Spacetime}\KillingTwoform = 2$,
the worldsheet is given by an integral manifold of $\ker \KillingTwoform$.

The equation of motion and the gauge conditions are generally reduced to
first order ordinary differential equations on a hypersurface
$\Hypersurface_0$ equipped with the $1$-form $\KillingOneform$. 
This $1$-form enables us to take a suitable coordinate
system on the hypersurface $\Hypersurface_0$,
and then it is shown that the equations are integrable.

The metric dual $1$-form $\KillingOneform$ provides the hypersurface
$\Hypersurface_0$ with an almost contact structure.
In the special case that $\corank_{\Hypersurface_0} \KillingTwoform = 1$,
the almost contact structure becomes a contact structure,
and its Reeb vector field gives the solutions to the ordinary differential
equations to be solved.
That is to say, the worldsheets are completely characterized by the Reeb
vector field.

We have also applied our formalism to some four-dimensional spacetimes:
{\Ppwaves} in which $\KillingTwoform = 0$, and 
the Einstein static universe and the G\"{o}del universe in which 
$\KillingTwoform \neq 0$.
The string worldsheets are obtained exactly and their geometries are
investigated in detail.

Our work complements previous studies of {\COne} string dynamics,
where the {\COne} symmetry was implicitly assumed to be non-null.
It shows that a null {\COne} symmetry is special
in the sense that the equation of motion is always integrable.
For strings with a non-null {\COne} symmetry,
the integrability requires additional spacetime symmetries
such as Killing vector fields and Killing tensor fields.
This point is one of the remarkable differences between null and non-null
{\COne} symmetries.

The concept of the cohomogeneity one symmetry is extended to
higher dimensional objects such as membranes
\cite{Kozaki:2014aaa,Hasegawa:2018cyk}.
The application of the null cohomogeneity one symmetry to the higher dimensional objects
is left for future work.

Our study reveals the existence of (almost) contact structure in the curved
spacetimes that admit a null Killing vector field and its relation to the
string dynamics. 
Applications of the (almost) contact structure to general relativity,
such as the construction of solutions to the Einstein equations,
may be intriguing.

\begin{acknowledgments}
      We are grateful to Osaka Central Advanced Mathematical Institute:
      MEXT Joint Usage/Research Center on Mathematics and Theoretical
      Physics JPMXP0619217849.
      TK acknowledges JSPS KAKENHI Grant Number JP20K03772 and 
      MEXT Quantum Leap Flagship Program (MEXT Q-LEAP) Grant Number
      JPMXS0118067285.
\end{acknowledgments}

\appendix
\section{Almost contact structure}
\label{appendix:almost_contact_structure}
We provide an overview of the almost contact structure and related topics 
\cite{Sasaki:1960,Blair:1976}.

An \emph{almost contact structure} on a $(2\IntOdd + 1)$-dimensional
manifold is characterized by a triplet
$\AlmostContactStructure$, 
where $\Endmorphism$ is a $(1,1)$-tensor, $\xi$ a vector field and
$\Oneform$ a $1$-form, such that
\begin{align}
  \iota_{\VectorField} \Oneform
  =
  \Oneform(\VectorField)
  = 1,
  \label{eq:condition_almost_contact1}
  \\
  \Endmorphism^2
  =
  -1 + \VectorField \otimes \Oneform.
  \label{eq:condition_almost_contact2}
\end{align}
It is readily shown that
\begin{align}
  \Endmorphism \, \VectorField
  = 0,
  \quad
  \Oneform \circ \Endmorphism = 0,
  \quad
  \rank(\Endmorphism) = 2 \IntOdd.
  \label{eq:propisition_almost_contact}
\end{align}

We can show that any odd-dimensional manifold with a nonzero $1$-form
$\Oneform$ admits an almost contact structure
$(\Endmorphism, \VectorField, \Oneform)$,
that is,
we can find $\VectorField, \Endmorphism$ satisfying
Eqs.~\eqref{eq:condition_almost_contact1} and
\eqref{eq:condition_almost_contact2} for a given $\Oneform$.
First we take a vector field $\VectorField$ that satisfies
Eq.~\eqref{eq:condition_almost_contact1}.
We note that the choice is not unique.
Next we take $2\IntOdd$ independent vector fields
$\VectorField_1,\dots,\VectorField_{2\IntOdd}$
such that $\Oneform(\VectorField_{i}) = 0 ~(i = 1,\dots,2\IntOdd)$.
Then we define the $(1,1)$-tensor $\Endmorphism$
so that 
\begin{align}
  \Endmorphism \,\VectorField
  =
  0,
  \quad
  \Endmorphism \, \VectorField_{2k-1}
  =
  \VectorField_{2k},
  \quad
  \Endmorphism \, \VectorField_{2k}
  =
  - \VectorField_{2k-1},
  \label{eq:action_of_endmorphism}
\end{align}
where $k = 1,\dots,\IntOdd$.
In this manner, we obtain an almost contact structure $\AlmostContactStructure$.

It is known that a manifold with an almost contact structure
$\AlmostContactStructure$
admits a Riemannian 
\emph{compatible} metric such that 
\begin{align}
  \Metric(\Endmorphism V_1, \Endmorphism V_2)
  =
  \Metric(V_1, V_2) - \Oneform(V_1) \,\Oneform(V_2)
  \label{eq:condition_almost_contact_metric}
\end{align}
for any vector fields $V_1$ and $V_2$.
Substituting $V_1 = \VectorField$,
we see that $\Oneform$ and $\VectorField$ are dual with respect to the
compatible metric 
\begin{align}
  \Oneform (V_2)
  =
  \Metric(\VectorField, V_2).
\end{align}
Furthermore, the norm of $\VectorField$ is unity;
$\Metric(\VectorField, \VectorField) = \Oneform(\VectorField) = 1$.
The compatible metric is generalized to the Lorentzian signature
by replacing Eq.~\eqref{eq:condition_almost_contact_metric} with
\begin{align}
  \Metric(\Endmorphism V_1, \Endmorphism V_2)
  =
  \Metric(V_1, V_2)
  -
  \epsilon \Oneform(V_1) \,\Oneform(V_2).
  \label{eq:condition_almost_contact_metric_Lorentz}
\end{align}
where $\epsilon = \pm 1$ \cite{Calvaruso:2010}.
An almost contact manifold with a compatible metric $\Metric$ is said
to have an \emph{almost contact metric structure} $\AlmostContactMetricStructure$.

If the compatible metric $\Metric$ satisfies
\begin{align}
  \Metric(V_1, \Endmorphism V_2)
  =
  \dd{\Oneform}(V_1,V_2)
\end{align}
for any vector fields $V_1$ and $V_2$,
the almost contact metric structure $\AlmostContactMetricStructure$
is called a \emph{contact metric structure}.
In this case, it holds that 
\begin{align}
  \Oneform \wedge \qty(\dd{\Oneform})^{\IntOdd}
  \neq  0,
  \label{eq:condition_contact}
\end{align}
where
\begin{align}
  \qty(\dd{\Oneform})^{\IntOdd}
  \coloneqq
  \underbrace{\dd{\Oneform} \wedge \dots \wedge
  \dd{\Oneform}}_{\text{$\IntOdd$ factors}}.
\end{align}

Conversely, a $(2\IntOdd + 1)$-dimensional manifold furnished with a
$1$-form $\Oneform$ satisfying Eq.~\eqref{eq:condition_contact}
is said to have a \emph{contact structure}.
A contact manifold admits a unique vector field
$\VectorField$ that satisfies 
\begin{align}
  \iota_{\VectorField} \Oneform = 1,
  \quad
  \iota_{\VectorField} \dd{\Oneform} = 0.
  \label{eq:condition_reeb}
\end{align}
This vector field $\VectorField$ is called the Reeb vector field.
Hereafter, for a contact manifold,
we only consider the almost contact structure $\AlmostContactStructure$
and the contact metric structure $\AlmostContactMetricStructure$
such that $\VectorField$ is the Reeb vector field.
The contact metric structure $\AlmostContactMetricStructure$ is called a \emph{$K$-contact structure}
if the Reeb vector field $\VectorField$ is a Killing vector
with respect to $\Metric$.

An almost contact structure $\AlmostContactStructure$ is said to be
\emph{normal} if
\begin{align}
  \nijenhuis_{\Endmorphism}
  +
  2 \VectorField \otimes \dd{\Oneform}
  =
  0
\end{align}
holds,
where $\nijenhuis_{\Endmorphism}$ is a $(1,2)$-tensor called the Nijenhuis
tensor defined by
\begin{align}
  \nijenhuis_{\Endmorphism}(V_1,V_2)
  \coloneqq
  \Endmorphism^{2} \, \qty[V_1, V_2]
  +
  \qty[\Endmorphism \, V_1, \Endmorphism \, V_2]
  -
  \Endmorphism \, \qty[\Endmorphism V_1, V_2]
  -
  \Endmorphism \, \qty[V_1, \Endmorphism V_2]
\end{align}
for any vector fields $V_1$ and $V_2$.
If the contact metric structure $\AlmostContactMetricStructure$ is normal,
the manifold is said to be a \emph{Sasakian manifold}.
We remark that there exist other equivalent definitions of the Sasakian
manifold.

Finally, we note that
for a given almost contact metric structure
$(\Endmorphism, \VectorField, \Oneform, h)$,
the following
$(\tilde{\Endmorphism}, \tilde{\VectorField}, \tilde{\Oneform},
\tilde{h})$ is also an almost contact metric structure
\begin{align}
  \tilde{\Endmorphism} = \Endmorphism,
  \quad
  \tilde{\VectorField} = (\lambda + \mu)^{-1/2} \VectorField,
  \quad
  \tilde{\Oneform} = (\lambda + \mu)^{1/2} \Oneform,
  \quad
  \tilde{\Metric} = \lambda h + \mu \Oneform \otimes \Oneform,
  \label{eq:transformation_acstructure}
\end{align}
where $\lambda, \mu$ are functions such that
$\lambda > 0, \lambda + \mu > 0$ \cite{Alexiev:1986}.

\section{Second fundamental form and twist potential}
\label{appendix:second_fundamental_form}
We provide an overview of the mathematical description of the worldsheet 
$\Worldsheet$ viewed as a two-dimensional submanifold embedded in a
$(\SpacetimeDimension)$-dimensional spacetime $\Spacetime$
\cite{Kobayashi:1996kbs,Capovilla:1994bs}.
The codimension of $\Worldsheet$ is denoted by $\Codimension$.

Let $\SetVecField(\Worldsheet)$ be the set of all tangent vector fields on
$\Worldsheet$ and $\SetVecField(\Worldsheet)^{\perp}$ be that of normal
vector fields on $\Worldsheet$.
The \emph{second fundamental form} $\SecFundForm$ is a symmetric map
\begin{align}
  \SecFundForm
  :
  \SetVecField(\Worldsheet) \times \SetVecField(\Worldsheet)
  \to
  \SetVecField(\Worldsheet)^{\perp}
\end{align}
such that for $X,Y \in \SetVecField(\Worldsheet)$ and
$p \in \Worldsheet$,
\begin{align}
  \bigl .\SecFundForm(X,Y) \bigr|_{p}
  =
  \qty(
  \nabla_{X} Y
  )_{p}^{\perp},
\end{align}
where $\perp$ denotes the projection to the normal complement
of $T_{p}\Worldsheet$ in $T_{p}\Spacetime$.
Then it holds that for $N \in \SetVecField(\Worldsheet)^{\perp}$
\begin{align}
  \Bigl. g(N, \SecFundForm(X,Y)) \Bigr|_p
  =
  \Bigl. g(N, \nabla_X Y) \Bigr|_p.
  \label{eq:ip_secfund_normalvec}
\end{align}
Let $N_{I} ~(I = 1,\dots, \Codimension)$ be normal vector
fields which are independent at each point on $\Worldsheet$.
Then we express $\SecFundForm(X,Y)$ by
\begin{align}
  \SecFundForm(X,Y)
  =
  K^{I}(X,Y)\, N_I,
  \label{eq:expansion_SecFundForm}
\end{align}
where $K^{I}~(I = 1, \dots, \Codimension)$ are symmetric maps
from $\SetVecField(\Worldsheet) \times \SetVecField(\Worldsheet)$ to
$\SetFunc(\Worldsheet)$~
($\SetFunc({\Worldsheet})$ being the set of all functions on 
$\Worldsheet$).
Substituting Eq.~\eqref{eq:expansion_SecFundForm} to
Eq.~\eqref{eq:ip_secfund_normalvec} with $N = N_J$,
we have
\begin{align}
  g_{JI}K^{I}(X,Y)
  =
  g(N_J, \nabla_X Y),
\end{align}
where $g_{JI} \coloneqq  g(N_J, N_I)$,
and then, the symmetric maps $K^I$ are given by
\begin{align}
  K^{I}(X,Y)
  =
  g^{IJ} g(N_J, \nabla_X Y),
\end{align}
where $g^{IJ}$ is the inverse of $g_{IJ}$.
It is more convenient to consider the symmetric map
$K_I \coloneqq g_{IJ}K^J$ such that
\begin{align}
  K_I(X,Y)
  =
  g(N_I, \nabla_X Y).
  \label{eq:SecondFundamentalForm}
\end{align}
Let $\zeta^a~(a = 1,2)$ be coordinates on the worldsheet $\Worldsheet$,
then the coordinate components $\qty(K_{I})_{ab}$ are given by
\begin{align}
  \qty(K_{I})_{ab}
  \coloneqq
  K_{I}(\pdv{}{\zeta^a}, \pdv{}{\zeta^b})
  =
  \qty(N_I)_{\mu}
  \qty[
  \pdv{}{\zeta^a}
  \qty(
  \pdv{x^{\mu}}{\zeta^b}
  )
  +
  \Gamma^{\mu}{}_{\nu\lambda}
  \pdv{x^{\nu}}{\zeta^a}
  \pdv{x^{\lambda}}{\zeta^b}
  ].
\end{align}

Suppose that a Killing vector field $\KillingVector$
of constant norm is tangent to the worldsheet.
As shown in Subsec.~\ref{subsec:c1_string_null},
$\KillingVector$ satisfies the geodesic equation
$\nabla_\KillingVector \KillingVector = 0$.
Therefore, it holds that $K_I(\KillingVector, \KillingVector) = 0$.
Taking one of the worldsheet coordinate, say $\zeta^1$,
so that $\pdv*{}{\zeta^1} = \KillingVector$,
we obtain $\qty(K_{I})_{11} = 0$.

In terms of the symmetric maps $K_{I}$,
the Nambu-Goto equation \eqref{eq:NambuGotoEqGeneral}
reduce to 
\begin{align}
  \Tr K_I
  \coloneqq
  \gamma^{ab} \qty(K_I)_{ab}
  =0,
\end{align}
where $\gamma_{ab}$ is the induced metric on $\Worldsheet$. 
In fact, when we  write Eq.~\eqref{eq:NambuGotoEqGeneral} as
\begin{align}
  \pdv{}{\zeta^a}
  \qty(
  \sqrt{-\gamma} \, \gamma^{ab}
  )\,
  \pdv{x^{\mu}}{\zeta^b}
  +
  \sqrt{-\gamma} \, \gamma^{ab}
  \qty[
  \pdv{}{\zeta^a}
  \qty(
  \pdv{x^{\mu}}{\zeta^b}
  )
  +
  \Gamma^{\mu}{}_{\nu\lambda}
  \pdv{x^{\nu}}{\zeta^a}
  \pdv{x^{\lambda}}{\zeta^b}
  ]
  =
  0
\label{eq:NG_eq_general_K}
\end{align}
and take the inner products with
the normal vector fields $N_{I}$,
we have
\begin{align}
  \gamma^{ab}
  \qty(N_{I})_{\mu}
  \qty[
  \pdv{}{\zeta^a}
  \qty(
  \pdv{x^{\mu}}{\zeta^b}
  )
  +
  \Gamma^{\mu}{}_{\nu\lambda}
  \pdv{x^{\nu}}{\zeta^a}
  \pdv{x^{\lambda}}{\zeta^b}
  ]
  =
  \gamma^{ab}
  \qty(K_I)_{ab}
  =0.
\end{align}

In order to define a twist potential,
we consider a map
\begin{align}
  \beta : \SetVecField(\Worldsheet) \times \SetVecField(\Worldsheet)^{\perp}
  \to \SetVecField(\Worldsheet)^{\perp}
\end{align}
such that for $X \in \SetVecField(\Worldsheet)$,
$N \in \SetVecField(\Worldsheet)^{\perp}$ and $p \in \Worldsheet$
\begin{align}
  \Bigl. \beta (X,N) \Bigr|_{p}
  =
  \qty(\nabla_{X} N)^{\perp}_{p}.
\end{align}
For a given set of independent normal vector fields $\qty{N_I}$, we  express $\beta$ by
\begin{align}
  \beta(X,N_I)
  =
  \TwistPotential_I{}^{J}(X) N_J,
\end{align}
where $\TwistPotential_{I}{}^{J} \quad (I,J = 1,\dots,\HypersurfaceDimension - 1)$ are
maps from $\SetVecField(\Worldsheet)$ to $\SetFunc(\Worldsheet)$,
that is, $1$-forms on $\Worldsheet$,
which are given by 
\begin{align}
  \TwistPotential_I{}^J (X)
  =
  g(\nabla_X N_I, N_K) \, g^{KJ}.
  \label{eq:definition_twist_potential}
\end{align}
When we define $1$-forms $\TwistPotential_{IJ}$ as
\begin{align}
  \TwistPotential_{IJ}
  \coloneqq
  g_{JJ'}\TwistPotential_I{}^{J'},
\end{align}
we can show that for any $X \in \SetVecField(\Worldsheet)$
\begin{align}
  \TwistPotential_{IJ}(X) + \TwistPotential_{JI}(X)
  =
  X (g_{IJ}).
  \label{eq:AntiSymmetryOmega}
\end{align}

When we take another set of independent normal vector fields
$G_I{}^J N_J \quad (G_I{}^J \in \mathrm{GL}(\HypersurfaceDimension - 1))$,
the maps $\TwistPotential_I{}^J$ are transformed as
\begin{align}
  \qty(\TwistPotential_I{}^J)_a
  \mapsto
  G_I{}^{I'}\,
  \qty(\TwistPotential_{I'}{}^{J'})_a
  \qty(G^{-1})_{J'}{}^J
  +
  \pdv{}{\zeta^a}\qty(G_I{}^{I'}) \qty(G^{-1})_{I'}{}^{J},
\end{align}
where $\qty(\TwistPotential_I{}^J)_a$ are the coordinate components 
\begin{align}
  \qty(\TwistPotential_I{}^J)_a
  \coloneqq
  \TwistPotential_I{}^J (\pdv{}{\zeta^a})
  =
  g_{\mu\nu}
  \qty[
  \pdv{\qty(N_I)^{\mu}}{\zeta^a}
  +
  \Gamma^{\mu}{}_{\alpha\beta} \, \pdv{x^{\alpha}}{\zeta^a} \qty(N_I)^{\beta}
  ]
  \qty(N_K)^{\nu} g^{KJ}.
  \label{eq:twist_potential_formula}
\end{align}
Since $\qty(\TwistPotential_I{}^J)_a$ transforms as a connection,
we can define the curvature $2$-forms $\Omega_I{}^J$ associated with 
$\TwistPotential_I{}^J$ as
\begin{align}
  \Omega_I{}^J
  =
  \dd{\TwistPotential}_I{}^J
  +
  \TwistPotential_I{}^K \wedge \TwistPotential_K{}^J.
\end{align}
The coordinate components are given by
\begin{align}
  \qty(\Omega_I{}^J)_{ab}
  \coloneqq
  D_{a} \qty(\TwistPotential_I{}^J)_b
  -
  D_{b} \qty(\TwistPotential_I{}^J)_a
  +
  \qty(\TwistPotential_I{}^K)_a \qty(\TwistPotential_K{}^J)_b
  -
  \qty(\TwistPotential_I{}^K)_b \qty(\TwistPotential_K{}^J)_a,
\end{align}
where $D_a$ denotes the covariant derivative on $\Worldsheet$.

When we impose independent normal vector fields $N_I$ to be orthonormal,
that is
$g_{IJ} = g(N_I, N_J) = \delta_{IJ}$,
it follows from Eq.~\eqref{eq:AntiSymmetryOmega} that 
$\TwistPotential_{IJ}$ are antisymmetric with respect to the indices $I,J$,
and the maps $\TwistPotential_I{}^{J}$ are called the
\emph{(extrinsic) twist potential}.

\section{Derivation of Eq.~\eqref{eq:NonTrivialComponentsKandProjection}}
\label{appendix:derivation}

The orthogonal projection $P_n : T_p \Spacetime \to T_p\Hypersurface_0$
is given by
\begin{align}
  \qty(P_n)^\mu{}_\nu = \delta^\mu{}_\nu - \epsilon \,n^\mu n_\nu,
  \quad
  \epsilon
  =
  \begin{cases}
    1 & \text{for $g(n,n) = 1$}\\
    -1 & \text{for $g(n,n) = -1$}
  \end{cases}.
\end{align}
Then a vector $X \in T_p \Spacetime$ is decomposed as
\begin{align}
  X = \epsilon g(n,X) n + P_n X.
  \label{eq:DecompositionOfX}
\end{align}
The null Killing vector $k$ is also decomposed as
\begin{align}
  k = \epsilon g(n,k) n + P_n k
  \label{eq:DecompositionOfKilling}
\end{align}
and hence, the unit vector $n$ perpendicular to the hypersurface
$\Hypersurface_0$ is given by
\begin{align}
  n = \frac{1}{\epsilon g(n,k)} \qty(k - P_n k ).
\end{align}
Substituting this equation into Eq.~\eqref{eq:DecompositionOfX},
we have
\begin{align}
  X 
  =
  \frac{g(n,X)}{g(n,k)} k + P_n X - \frac{g(n,X)}{g(n,k)} P_n k.
  \label{eq:kDecompositionOfX}
\end{align}
This equation gives the projection
$P_\KillingVector: T_p \Spacetime \to T_p \Hypersurface_0$
along $\KillingVector$ so that 
\begin{align}
  P_k X = P_n X - \frac{g(n,X)}{g(n,k)} P_n k
  \quad
  \text{or}
  \quad
  P_k X = X - \frac{g(n,X)}{g(n,k)} k.
  \label{eq:Projection}
\end{align}
Eqs.~\eqref{eq:DecompositionOfKilling} and \eqref{eq:Projection},
give the following formula that plays an important role in deriving 
Eq.~\eqref{eq:NonTrivialComponentsKandProjection}
\begin{align}
  g(P_k X, P_k Y) = g(P_k X, P_n Y)
  \quad
  \text{for $X, Y \in T_p \Spacetime, \quad g(X,k) = 0$}.
  \label{eq:Formula}
\end{align}

Now we derive Eq.~\eqref{eq:NonTrivialComponentsKandProjection}.
First we observe that,
for $\hat{l}\coloneqq P_k l$, 
\begin{align}
  K_I(\hat{l}, \hat{l})
  =
  K_I(l,l)
  \eqqcolon 
  \qty(K_I)_{\sigma\sigma},
\end{align}
where we have used Eq.~\eqref{eq:VanishingK1} and \eqref{eq:VanishingK2}.
Then, from Eq.~\eqref{eq:ComponentsOfSecondFundamentalForm},
$\qty(K_I)_{\sigma\sigma}$ is given by 
\begin{align}
  \qty(K_I)_{\sigma\sigma}
  =
  g(N_I, \nabla_{\hat{l}}\,\hat{l}\,).
  \label{eq:KwithPl}
\end{align}
Next, we decompose $N_I$ and $\nabla_{\hat{l}} \, \hat{l}$ by
using Eq.~\eqref{eq:Projection}.
Then Eq.~\eqref{eq:KwithPl} leads to
\begin{align}
  \qty(K_I)_{\sigma\sigma}
  =
  g(P_k N_I, P_k \nabla_{\hat{l}} \, \hat{l}).
  \label{eq:derivation}
\end{align}
In the process of the derivation, we have used
the equations $\nabla_k k = 0$, $\nabla_l k = 0$ and $g(k,l) = 1$,
which are different expressions of
Eqs.~\eqref{eq:GeodesicEqWithKillingOneform},
\eqref{eq:EOMC1string} and \eqref{eq:CoordinateConditionZeta2}
respectively.
Finally, using the formula \eqref{eq:Formula},
we obtain
\begin{align}
  \qty(K_I)_{\sigma\sigma}
  =
  g(P_k N_I, P_n \nabla_{\hat{l}} \, \hat{l}\,).
\end{align}

\bibliography{references}

\end{document}